\documentclass[a4paper,fleqn, 11pt]{cas-sc} 

\usepackage[numbers]{natbib}
\usepackage{lineno}
\usepackage[export]{adjustbox}[2011/08/13]
\usepackage{subfig}

\usepackage{float}
\floatstyle{plaintop}
\restylefloat{table}

\usepackage{url} 
\usepackage{tabularx,booktabs}
\usepackage{algorithm} 
\usepackage[noend]{algpseudocode} 
\usepackage{amsmath} 
\usepackage{bbm} 
\usepackage{setspace} 
\usepackage{float}
\usepackage{amsmath,amssymb}
\usepackage{graphicx}
\usepackage{caption}
\graphicspath{ {./figs/} }
\usepackage[normalem]{ulem}
\useunder{\uline}{\ul}{}

\begin{document}
\newcommand\barbelow[1]{\stackunder[1.2pt]{$#1$}{\rule{.8ex}{.075ex}}}
\let\WriteBookmarks\relax
\def\floatpagepagefraction{1}
\shorttitle{}

\title [mode = title]{Quantifying the multi-scale and multi-resource impacts of large-scale adoption of renewable energy sources}

\author[1]{Elnaz Kabir}
\cormark[1]
\ead{ek574@cornell.edu}

\address[1]{Biological $\&$ Environmental Engineering Department, The Cornell University, Ithaca, NY 14853, United States}

\author[1]{Vivek Srikrishnan}
\ead{vs498@cornell.edu}

\author[2]{M. Vivienne Liu}
\ead{ml2589@cornell.edu}

\address[2]{Systems Engineering, Cornell University, Ithaca, NY, 14850, USA}

\author[1]{Scott Steinschneider}
\ead{ss3378@cornell.edu}

\author[1,2]{C. Lindsay Anderson}
\ead{cla28@cornell.edu}


\begin{abstract}
The variability and intermittency of renewable energy sources pose several challenges for power systems operations, including energy curtailment and price volatility. In power systems with considerable renewable sources, co-variability in renewable energy supply and electricity load can intensify these outcomes. In this study, we examine the impacts of renewable co-variability across multiple spatial and temporal scales on the New York State power system, which is undergoing a major transition toward increased renewable generation. We characterize the spatiotemporal co-variability of renewable energy-generating resources and electricity load and investigate the impact of climatic variability on electricity price volatility. We use an accurate, reduced-form representation of the New York power system, which integrates additional wind, and solar power resources to meet the state's energy targets through $2030$. Our results demonstrate that renewable energy resources can vary up to $17\%$ from the annual average, though combining different resources reduces the overall variation to $\approx 8\%$. On an hourly basis, renewable volatility is substantially greater and may vary up to $100\%$ above and below average. This results in a $9\%$ variation in annual average electricity prices and up to a $56\%$ variation in the frequency of price spikes. While yearly average price volatility is influenced mainly by hydropower availability, daily and hourly price volatility is influenced by solar and wind availability.
\end{abstract}

\begin{keywords}
Renewable energy co-variability \sep Price variability \sep Price spike \sep Wind energy \sep Solar PV energy \sep Hydro energy 
\end{keywords}

\maketitle
\begin{onehalfspacing} 

\section{Introduction} \label{intro}
Global greenhouse gas (GHG) emissions have to decrease to net-zero by mid-century in order to meet the Paris Agreement target limiting global temperature increase to $2^{\circ}$C \cite{masson2022global}. Because the electricity sector is a major contributor to GHG emissions \cite{IEA2022CO2emissions}, power system decarbonization through the adoption of renewable energy sources (RES) is a high priority. It also has the potential to facilitate the decarbonization of transportation and buildings, so it has progressed rapidly in recent years. The European Union aims for RES to provide at least $40\%$ of their overall energy mix by $2030$, while China has committed to generating $35\%$ of its electricity from renewables by the same time.  In $2021$, the United States (U.S.) re-joined the Paris Agreement and set an ambitious plan to reduce net GHG emissions by around $50\%$ below $2005$ levels by $2030$ \cite{USNetZeroPlan2021}, and to deliver net-zero emissions no later than $2050$. Accordingly, the U.S. plans to achieve $100\%$ clean electricity by $2035$ \cite{USNetZeroPlan2021}. 

As countries accelerate their transition to carbon-free energy systems, they must contend with several challenges associated with the penetration of renewable energy into the power system. Renewable energy sources are highly variable and intermittent \cite{alshawaf2020solar, ela2013impacts,naegele2020climatology}. This includes variability in peak values and consistency of renewable energy generation, co-variability between renewable sources and load across timescales \cite{alshawaf2020solar, naegele2020climatology}, and the spatial complementarity of renewable sources across regions \cite{ela2013impacts}. Even though the characterization of renewable energy variability has gained attention, much of the effort has concentrated on solar energy (e.g., \cite{alshawaf2020solar, ela2013impacts, abunima2022two}) and wind energy (e.g., \cite{naegele2020climatology, quan2019survey, pereira2019effect}), in isolation or considering co-variability across wind and solar \cite{bett2016climatological, monforti2014assessing}. Studies have, less commonly, included co-variability with hydropower (subsequently referred to as "hydro") and the drivers of electricity load (i.e., temperature), which can also vary over decadal time scales. Studies of renewable variability have often focused on certain regions (e.g., Europe \cite{monforti2014assessing, bett2016climatological} and Western U.S. \cite{mohammadi2018study}) and less frequently on other regions undergoing energy transitions. 
 
Increasing variable renewable energy penetration results in other challenges in the absence of sufficient storage and transmission capacity, such as increased energy curtailment, supply shortages, and transmission congestion. These challenges are exacerbated by difficulties in forecasting renewable energy production at intra-daily timescales \cite{quan2019survey} and operational strategies are needed to mitigate these additional uncertainties \cite{abunima2022two, zhao2021new}. Accurately quantifying these risks is instrumental for developing strategies to manage them, e.g., \cite{huang2021economic, bird2016wind}. There is a research gap in characterizing how operational constraints interact with substantial amounts of hydro, wind, and solar variability across various spatial and temporal scales. 

Another critical challenge renewable energy variability introduces is increased electricity price volatility \cite{pereira2019effect}. When price volatility rises, long-term investors may delay investment or increase risk management activities, raising electricity prices. Price volatility can be increased by changes in renewable energy curtailment, supply shortage, and transmission congestion, though stable regulatory policies can reduce these effects \cite{ciarreta2020renewable, pereira2019effect}. The impacts of variable renewable energy and the above bottlenecks on price volatility are region-specific \cite{rintamaki2017does} and may differ across time. Therefore, studying the implications of the above constraints on the price distribution across space and time scales is needed. 

To address the aforementioned research gaps, we explore the RES co-variability in a hybrid energy system with large-scale wind, solar, and hydropower deployment across a multi-decadal time period. We study the impacts of this co-variability and its interaction with the structure and operation of the New York State (NYS) power system, primarily  renewable energy curtailment and electricity price volatility. We also identify the meteorological and climatic conditions associated with increases to these risks. 
We use a representation of the NYS power system \cite{liu2022open},  modified to meet the state's 2030 decarbonization target \cite{BillS6599}. We develop a multi-period DC optimal power flow model which integrates RES and energy storage units into the system. Using a 40-year historical record of meteorological variables, we simulate hourly wind, solar, and hydropower across NYS and temperature-based electricity demand. Producing renewable energy and electricity load data for this long time horizon while preserving their spatiotemporal co-variability allows us to investigate renewable energy and electricity load co-variability and their consequent risks, including energy curtailment and price volatility, under a wide range of weather conditions. Finally, we do sensitivity analysis to evaluate the sensitivity of results to the central assumptions made in the model. 

The rest of this paper is structured as follows: Section \ref{NYpowerSystemBackground} introduces the NYS power grid and the NYS Climate Leadership and Community Protection Act (CLCPA) decarbonization plan as our case study. Section \ref{Model} describes the multi-period DC optimal power flow formulation used to model renewable resource integration in the power grid. It then summarizes the actual representation of the NYS power grid and the modifications made to meet the state's decarbonization targets. Section \ref{Data} explains how needed data is generated and integrated into the model, and section \ref{Results} represents all the numerical results. Section \ref{Discussion} discusses the sensitivity of results to model assumptions and possible improvements that can be made to the model and data. Finally, section \ref{Conclusion} summarizes and concludes the paper. 
\section{Background on the New York Power System} \label{NYpowerSystemBackground}
NYS has one of the most ambitious decarbonization targets among U.S. states \cite{CESA_cleanEnergyStates}. These goals are codified in the Community Leadership and Climate Protection Act (CLCPA) \cite{CLCPA}. The CLCPA promises to increase the RES share to $70\%$ of total electricity load by $2030$ and to have a zero-emission grid by $2040$. NYS is also one of the country's largest hydropower producers, allowing the state to combine hydropower with wind and solar energy sources and have an accelerated transition toward its clean energy target. Having (i) an ambitious renewable target and (ii) a large source of hydropower makes the NYS power grid an appropriate case study to understand the potential challenges and benefits of this transition.
.  

\begin{figure}[t]
\includegraphics[width=0.7\textwidth]{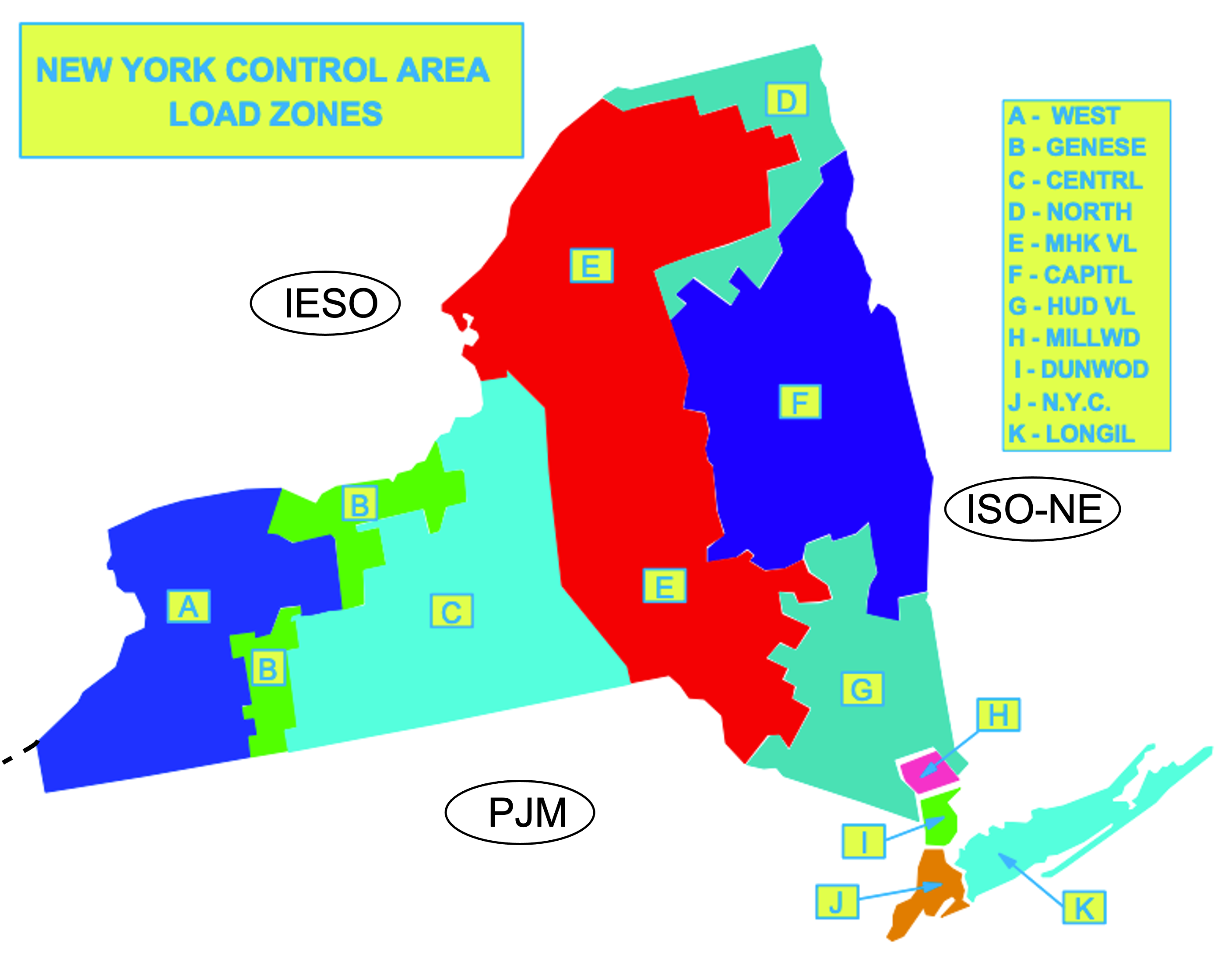}
\centering
\caption{Layout of the 11 internal and 3 external load zones in the New York Control Area (source: NYISO).}
\label{NYCA_11zones}
\end{figure}
Following the state's power system transition plan, the New York Independent System Operator (NYISO) commissioned several studies focusing on the state's long-term energy and peak load, power system reliability, and future capacity needs, establishing a framework for this paper's analysis. The study in \cite{NYISOphaseI2019} developed a long-term energy and peak load forecast model for the state's load zones (see Figure \ref{NYCA_11zones}), reflecting the potential impact of electrification of transportation and heating. Building on that study, the work in \cite{NYISOphaseII2020} simulated the potential impacts of climate change and climate policy on power system operations and provided updated energy and peak load forecasts. Every year, NYISO also releases a ``Gold Book" providing forecasted and realized peak demand and energy usage under different scenarios, including under the CLCPA. The 2021 Gold Book \cite{GoldBook2021} estimates energy use and peak demand considering projections of weather trends under climate change, economic growth, energy efficiency, behind-the-meter (BTM) solar photovoltaic (PV) and non-solar generators, energy storage, and electrification. 

To meet the CLCPA target for $2030$, the Congestion Assessment and Resource Integration Study \cite{CARIS2019} developed a zonal-level renewable capacity build-out scenario (i.e., the \textit{70$\times$30} scenario), whereby installed renewable energy equaled approximately 70$\%$ of the energy consumed in the state. A related effort \cite{RNA2020} evaluated the reliability of the New York bulk electric grid through 2030 based on planned upgrades to the transmission system, and in the process developed the topology of the transmission system, interface limits between zones, and transfer limits between the New York control area and the external areas for $2030$. The work presented in the current study builds on many of the system modeling assumptions used in these earlier analyses. 
 
\section{Model} \label{Model}
This paper, by applying a model predictive control strategy, develops the mathematical model using a multi-period DC optimal power flow formulation with storage units. The model provides an optimal dispatch and storage strategy while minimizing generation costs and satisfying all the electricity load. 

\subsection{Mathematical Formulation} \label{DCOPF}
We consider a power transmission system with a set of nodes/buses $\mathcal{N} = \{1, \ldots, N\}$ connected through a set of power flow lines $\mathcal{L} \subseteq \mathcal{N} \times \mathcal{N}$. Bus $k$ is connected to bus $m$ if and only if $(k,m) \in \mathcal{L}$. At the bus $k \in \mathcal{N}$, $\mathcal{N}(k)$ and $\mathcal{G}(k)$ denote the set of all connected buses and generators, respectively. $\mathcal{I}_k$ and $\mathcal{O}_k$ are, respectively, the set of lines flowing into bus $k$ and the set of lines flowing out of bus $k$. $E_k$ is the active power demand at the bus $k$, and we assume all electricity demands are independent of electricity price and known beforehand. We predict this demand according to historical temperature data. The NYS power system is divided into spatial zones (see figure \ref{NYCA_11zones}); we denote the set of interfaces between these zones by $\mathcal{I}$. Each interface  $i \in \mathcal{I}$  consists of a set of lines denoted by $\mathcal{IF}_i \subset \mathcal{L}$. We divide the time period into $T$ time periods $\mathcal{T}=\{1, \ldots, T\}$. Finally, $\mathcal{R} = \{H, S, W \}$ represents the set of renewable types, namely hydro, solar, and wind.

\textbf{Constraints}: The model includes constraints related to the curtailment of RES, power grid operation, charge/discharge of batteries, and electricity load, as explained below. 


\textit{Curtailment of renewable energy sources}:
Suppose the active power curtailment of RES at the node $k \in \mathcal{N}$ and time slot $t \in \mathcal{T}$ is $p_{k,r}(t)$. The following bound constraints are imposed:
\begin{align}
& 0 \leq p_{k,r}(t)\leq  \overline{P}_{k,r}(t) ,  &t \in \mathcal{T}, k \in \mathcal{N}, r \in \mathcal{R}\label{REScurtailConst}
\end{align}
where $\overline{P}_{k,r}(t)$ is the known RES maximum active power generation. 

\textit{Power grid operation}:
For each distributed generator $g \in \mathcal{G}(k)$ at bus $k$ and time $t$, the following constraints associated with the active generation $p_{k,g}(t)$ must be fulfilled:
\begin{align}
& \underline{P}_{k,g} \leq p_{k,g}(t)\leq  \overline{P}_{k,g} ,  								&t \in \mathcal{T}, k \in \mathcal{N}, g \in \mathcal{G}(k) \label{DGlimitConst}\\
& \underline{R}_{k,g} \leq p_{k,g}(t)  - p_{k,g}(t-1)  \leq \overline{R}_{k,g},   \text{   }	&t \in \mathcal{T}, k \in \mathcal{N}, g \in \mathcal{G}(k)\label{RamplimitConst} \\
&  \underline{L} \leq e_l(t) \leq \overline{L}, \text{   }										& t \in \mathcal{T}, l \in \mathcal{L}\label{FlowlimitConst} \\
&  \underline{L}_{IF_i} \leq \sum_{l \in IF_i}e_{l}(t) \leq \overline{L}_{IF_i}, \text{   }& t \in \mathcal{T}, i \in \mathcal{I}\label{IFlowlimitConst} \\
& e_{l}(t) = B_l (\theta_{k}(t) - \theta_{m}(t))\text{   }& t \in \mathcal{T}, l=(k,m) \in \mathcal{L}. \label{phaseAngleConst}
\end{align}

The constraints \eqref{DGlimitConst} and \eqref{RamplimitConst} are the active power bounds and ramping constraints of distributed generators with $\underline{P}_{k,g}$, $\overline{P}_{k,g}$ and $\underline{R}_{k,g}$, $\overline{R}_{k,g}$ denoted as the corresponding limitations. The constraint \eqref{FlowlimitConst} represents the bounds on the power flow magnitude $e_l(t)$ for each transmission line $l$, where $\underline{L}$ and $\overline{L}$ denote the corresponding lower and upper bounds on each transmission line. The constraints \eqref{IFlowlimitConst} limit the amount of power transferred between each pair of zones through a group of transmission lines connecting those zones. The constraint \eqref{phaseAngleConst} guarantees the flow balance in terms of the phase angle with $\theta_k(t)$ where $\theta_{k}(t)$ is the phase angle of bus $k$, and $B_l$ is the susceptance parameter of the transmission line $l$ connecting bus $k$ to $m$. 

\textit{Battery energy storage}:
At each bus $k$, we consider a single energy storage system that may be built of multiple batteries, though power does not flow between them. The state of charge dynamics $s_k(t)$ for the battery energy storage at bus $k$ is modeled using the following discrete-time state equations. 

\begin{align}
& s_k(t+1) = s_k(t)-(\sqrt{SE_{k}} psc_k(t) + \frac{1}{\sqrt{SE_{b}}} psd_k(t)) \text{   }& t \in \mathcal{T}, k \in \mathcal{N} \label{SEbalanceConst}\\
& S^{min}_{k} \leq s_{k}(t) \leq S^{max}_{k} \text{   }& t \in \mathcal{T}, k \in \mathcal{N} \label{SElimitConst}\\
& ps_k(t) = psc_k(t) + psd_k(t) \text{   }& t \in \mathcal{T}, k \in \mathcal{N} \label{SPbalanceConst}\\
& psd_k(t) - psc_k(t) \leq P_k \text{   }& t \in \mathcal{T}, k \in \mathcal{N} \label{SPlimitConst}\\
& psc_k(t) \leq 0 \;\; ; \;\; psd_{b,t} \geq 0 \text{   }& t \in \mathcal{T}, k \in \mathcal{N} \label{SEsignConst}
\end{align}
 
The constraints \eqref{SEbalanceConst} denote the increase/decrease in stored energy due to charging/discharging of the storage unit at bus $k$ during time $t$ while formulating the lost power due to charging/discharging batteries. $SE_{k}$ is the round-trip efficiency of the storage unit at the beginning of time $t$, and $psc_{k}(t)/psd_{k}(t)$ indicate the charge/discharge power injections of the storage unit during time $t$. The constraints \eqref{SElimitConst} are to assure that the state of charge of the battery (in MWh) is maintained within certain limits indicated as $S^{min}_k$ and $S^{max}_k$. Constraints \eqref{SPbalanceConst} calculate the amount of power injected to/absorbed from the power system at bus $k$ by the battery energy storage system. The constraints \eqref{SPlimitConst} denote that the charging power and discharging power (in MW) of the battery cannot go beyond the maximum power limit $P_{k}$.

\textit{Nodal power balance}: 
At any time slot $t$, the power supply and demand must be balanced for each bus $k$. Therefore, the DC power flow balance equation is:
\begin{align}
& \sum_{g \in G_{b}} p_{k,g}(t) + \sum_{r \in \mathcal{R}} \overline{P}_{k,r}(t) + \sum_{l \in I_k}e_{l}(t) + ps_{k}(t) + l_{k}(t)=\sum_{l \in O_k}e_{l}(t) + \sum_{r \in \mathcal{R}} p_{k,r}(t) + E_{k}(t)  &\text{   } t \in \mathcal{T}, k \in \mathcal{N}\label{eq: sec6}
\end{align}
where $p_{k,g}(t) + \overline{P}_{k,r}(t) + ps_k(t)$ is the power supplied by distributed generators, RES, and battery energy storage; $\sum_{l \in I_k}e_{l}(t)$ is the power transferred from other buses; $l_{k}(t)$ is the unmet electricity demand; $\sum_{l \in O_k}e_{l}(t)$ is the power transferred to other buses, $p_{k,r}(t)$ is the curtailed renewable power, and $E_{k}(t)$ is the electricity demand at bus $k$ during time $t$.

\textbf{Objective function}:
The primary objective of our multi-period DC optimal power flow model is to minimize the power generation costs and the unmet electricity demand while maximizing the dispatched RES. RES has  small/zero marginal costs which prioritize their dispatch compared to other energy sources. Thermal generator costs are assumed to be linear. A total cost-minimizing objective function is then:
\begin{align}
Min: \sum_{t \in \mathcal{T}}\sum_{k \in \mathcal{N}} \left( \sum_{g \in \mathcal{G}(k)}(C^0_{k,g}(t)+C^1_{k,g}(t)p_{k,g}(t)) + C^1_{k,H}(t)(\overline{P}_{k,H}(t) - p_{k,H}(t)) \right) \nonumber \\
\text{           } + M\sum_{t \in \mathcal{T}}\sum_{k \in \mathcal{N}} l_{k}(t) \label{Obj}
\end{align}
where the first term is the power generation cost and the second term is the total unmet electricity demand multiplied by the big $M$, a large cost for unserved energy. $C_{k,g}(t)$ and $C_{k,g}^{0}$ are, respectively, the linear and the constant cost coefficients of the generator $g$ at time $t$. $(\overline{P}_{k,H}(t) - p_{k,H}(t))$ is the amount of dispatched hydropower multiplied by the cost coefficient $C^1_{k,H}(t)$

\textbf{Locational marginal price}:
We formulate nodal electricity prices using locational marginal pricing (LMP). Electricity markets use LMPs to reflect the value of electric energy at different locations, accounting for the patterns of load, generation, and the physical limits of the transmission system. LMP is the change in electricity production cost (i.e., objective function (\ref{Obj})) to optimally deliver an increment of electricity load at the locations (i.e., $E_{k}(t) \text{    } \forall k,t$) while satisfying all the constraints. In our the DC-optimal power flow model, LMP is defined as the shadow price of the nodal power balance constraints (i.e., equation (\ref{eq: sec6})).

\subsection{Modified NYS power grid} \label{NYPowerGrid}
Our baseline grid model is the NYS power grid representation from \cite{liu2022open}, modified to represent the transmission system, renewable energy resources, and storage units suggested by the CLCPA plan (see Section \ref{Data}). The baseline grid model has 57 buses, and 94 transmission lines, where 46 buses are inside NYS and nine are in the neighboring areas. There are 227 thermal generators in the grid being modeled as dispatchable units. \cite{liu2022open} validated the model with both power flow and optimal power flow analyses using historical records from the NYISO.  

We modify the baseline NYS power grid to meet the CLCPA plan. The modifications include (i) updating the nuclear power generation according to their retirement plan, (ii) adding land-based and offshore wind farms, utility-based solar sites, and storage units, and (iii) updating interface flow capacity limits on the power flow between the 11 zones of NYS as well as the neighboring areas. These modifications are detailed in section \ref{Data}. 

\section{Data} \label{Data}
\begin{figure}[]
\includegraphics[width=\textwidth]{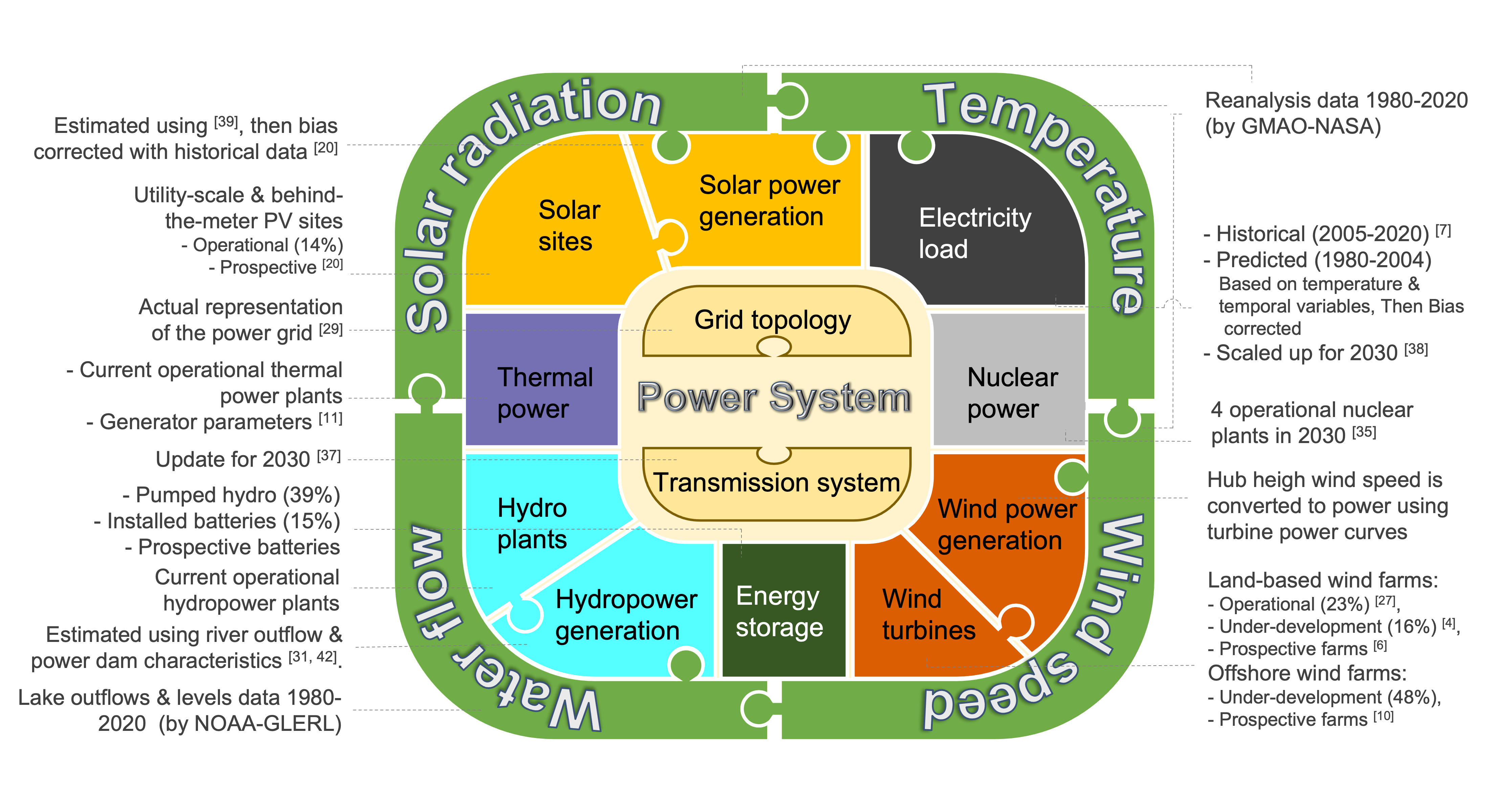}
\centering
\caption{Venn diagram of the developed power system with renewable power plants and battery storage integration considering co-variability in climate and environmental variables}
\label{Data_VenDiagram}
\end{figure}

 To characterize the co-variability in RES and load and their consequent risks while interacting with power system constraints, we gather influencing climate variables (i.e., solar radiation, temperature, and wind speed) and river environmental dynamics across NYS between 1980-2020. For each year of records, we estimate the renewable power generation of each power plant and the electricity load at each node. Using this data, we run the developed DC-OPF model proposed in Section \ref{DCOPF}. This provides operational attributes of the NYS power system under 40 samples of climatological and environmental scenarios in 2030. Figure \ref{Data_VenDiagram} represents the developed modeling framework. Estimating renewable generations and electricity demand in hourly temporal resolution and based on weather and environmental variables let us model the co-variability between these resources. Moreover, adopting the actual representation of the NYS power system while considering its operational details let us model the interaction of renewable generation and load with power grid complexities. In the following, we explain how we generate the forcings for these inputs and integrate these resources into the model.

\subsection{Wind power} \label{WindData}
To reach the state's target of $8761$ MW of land-based and $9000$ MW of offshore wind capacity, we add extra wind farms to the current wind generation footprint. The land-based wind farms include all operational and under-development wind farms as indicated by \cite{Hoen2018united} and \cite{NYSERDA_RenewableProjects}, respectively. To locate additional wind generation which is not currently under development \cite{NYSERDA_RenewableProjects}, we use the National Renewable Energy Lab (NREL) Wind Integration National Dataset (WIND) Toolkit (WTK) \cite{NRELWTK}, which provides a ranking of prospective locations for wind farms based on geography and wind resource statistics. From the $2859$ wind farms determined for NYS, we randomly choose farms  so that the overall capacity of wind farms in each zone satisfies the zonal capacity targets described in Table \ref{zonalCaps}. The offshore wind farms include the five under-development farms identified by \cite{nyserda_offshoreWind}, totaling $4300$ (MW) in capacity. Finally, to provide the unplanned offshore wind power capacity that remains, we model four additional offshore wind farms according to the study done in \cite{nyserda_offshoreWind}. To simulate the wind power generation, we use the 10m surface-level wind speed from the MERRA-2 reanalysis product between $1980-2019$. For each land-based and offshore wind farm, we interpolate the wind speed data to the 100m level and bias-correct it against the actual NREL WTK wind speed data, as in \cite{doering2023evaluating}. Then, we calculate wind power generation using the approach presented in \cite{draxl2015wind}. 

\subsection{Solar power} \label{SolarData}
We simulate and add 59 distributed-scale PV (DPV) and 62 utility-scale PV (UPV) solar sites into the grid. Their locations are determined according to NREL Solar Integration National Dataset (SIND) \cite{bloom2016eastern}; however, the proposed capacities in \cite{bloom2016eastern} are scaled to match the zonal capacities indicated in Table \ref{zonalCaps}. We add $3$ additional DPV solar sites near populated areas in zone D, J, and K, in order to meet the zonal capacities of DPV solar in Table \ref{zonalCaps}. To simulate the solar power generation at each solar site, we use temperature and incident shortwave radiation from the MERRA-2 reanalysis data between $1980-2019$. The method of statistical moments presented in \cite{perpinan2007calculation} is then employed to estimate solar power generation. Finally, the calculated values are bias-corrected using the actual NREL data. It should be noted that the DPV solar power generation is modeled as a negative load, while the UPV solar power generation is semi-dispatchable. 

\begin{table}[!b]
\centering
\caption{Zonal capacity (MW) of renewable and battery storage sources. }
\begin{tabular}{|l|lllllllllll|l|}
\hline
                    & \multicolumn{11}{c|}{{\ul Zone}}                                         &            \\
Technology Type     & A     & B   & C    & D    & E    & F    & G    & H   & I   & J    & K    & Total \\ \hline
Land-based wind     & 2692  & 390 & 1923 & 1935 & 1821 &      &      &     &     &      &      & 8761       \\
Offshore wind       &       &     &      &      &      &      &      &     &     & 6391 & 2609 & 9000       \\
UPV solar & 5748  & 656 & 3585 &      & 2268 & 4661 & 2636 &     &     &      & 77   & 19631      \\
DPV solar           & 704   & 218 & 596  & 69   & 673  & 827  & 684  & 61  & 90  & 672  & 846  & 5439       \\
Hydro               & 2,675 & 64  & 109  & 915  & 376  & 270  & 76   &     &     &      &      & 4485       \\
Battery storage     & 150   & 90  & 120  & 180  & 120  & 240  & 100  & 100 & 100 & 1320 & 480  & 3000       \\ \hline
\end{tabular}
\label{zonalCaps}
\end{table}

\subsection{Hydropower}\label{HydroData}
We model $10$ hydropower generators into the NYS power system. Two plants represent Robert Moses Niagara, with a capacity of $2675$ MW, and Moses-Saunders, with a capacity of $856$ MW. The remaining eight plants with an overall capacity of $953.7$ MW represent the existing 345 small hydro plants at an aggregated level. The quarter-monthly power generations at Moses-Niagara and the NYS share of Moses-Saunders dam are calculated using the procedures described in \cite{meyer2017evaluating} and \cite{semmendinger2022establishing}, respectively. To estimate the monthly generation of the other eight hydro plants, we calculate their aggregate monthly generation and their individual capacities, using NYISO fuel mix data \cite{NYISOfuelMix} and the Gold Book \cite{GoldBook2019}, respectively. We model the hydropower generation of eight small plants as a negative load, while generations of the two large hydro plants are treated as dispatchable. 

\subsection{Storage units} \label{StorageData}
To decrease congestion and increase utilization of renewable generation, the CLCPA supports the installation of $3000$ MW of storage units across NYS by $2030$. To represent this goal, we incorporate a storage unit at each bus, with each storage unit representing several smaller-sized batteries connected to the same bus. The total capacity of storage units per zone is summarized in Table \ref{storageCap} suggested by \cite{CARIS2019}. We disaggregate these zonal capacities down to the bus level such that the storage capacity in each bus be proportional to the renewable generation at that bus. This strategy results in battery sizes varying between 6 and 480 MW. We also model and include the Gilboa pumped hydro as a $1170$ MW storage unit. According to \cite{CARIS2019}, we assume the storage unit associated with the Gilboa pumped hydro has 12 hours of storage duration and round trip efficiency of $75\%$, while all other batteries have 8 hours of storage duration and operate with an 85$\%$ round trip efficiency. Finally, a minor marginal cost of 1$\$$/MW is assigned to the charging and discharging of batteries to encourage their use while avoiding a simultaneous charge and discharge.
\subsection{Electricity load} \label{EnergyLoadData}
To estimate hourly load in each NYS zone, we develop a random forest (RF) model using actual hourly zonal load data between $2005-2019$ from NYISO \cite{NYISOload}. The model's predictors include the day of the week, the day of the year, and the last 24 hours of average temperature throughout each zone. To guarantee that the variance of predictions is not underestimated by the model, we bias-correct the predicted loads using the quantile mapping approach. To construct a complete record of hourly zonal loads between $1980-2019$, we combine the RF model's predictions for $1980-2004$ with the available actual data for $2005-2019$. The combined load data is then scaled up to reach the summer and winter peak loads forecasted for $2030$ under the CLCPA case by NYISO \cite{GoldBook2021}. 
Table \ref{CLCPApeakLoad} compares the summer and winter peak loads of the entire NYS in years $2020$ and $2030$. 
Finally, zonal hourly load data are disaggregated and divided between buses according to their original load ratio in the NPCC-140 model \cite{chow1991user}.

\begin{table}[!b]
\centering
\caption{Assumed energy storage zonal power capacity} \label{storageCap}
\begin{tabular}{|lcccccccccccc|}
\hline
\multicolumn{13}{|c|}{Nameplate capacity distribution (MW)}              \\ \hline
& A & B & C & D & E & F & G & H & I & J & K & Total \\ \hline
\multicolumn{1}{|l|}{\begin{tabular}[c]{@{}l@{}}Storage \\ Capacity\end{tabular}} & \multicolumn{1}{c|}{150} & \multicolumn{1}{c|}{90} & \multicolumn{1}{c|}{120} & \multicolumn{1}{c|}{180} & \multicolumn{1}{c|}{120} & \multicolumn{1}{c|}{240} & \multicolumn{1}{c|}{100} & \multicolumn{1}{c|}{100} & \multicolumn{1}{c|}{100} & \multicolumn{1}{c|}{1320} & \multicolumn{1}{c|}{480} & 3000  \\ \hline
\end{tabular}
\end{table}
\begin{table}[!b]
\centering
\caption{NYS CLCPA-case peak load (MW) in years 2020 and 2030 \cite{GoldBook2021} }
\begin{tabular}{|l|cccc|}
\hline
               & \multicolumn{1}{c|}{\begin{tabular}[c]{@{}c@{}} summer peak load\\ - 2020 (Actual)\end{tabular}} & \multicolumn{1}{c|}{\begin{tabular}[c]{@{}c@{}} summer peak load\\ - 2030 (Forecast)\end{tabular}} & \multicolumn{1}{c|}{\begin{tabular}[c]{@{}c@{}}winter peak load\\ - 2020 (Actual)\end{tabular}} & \begin{tabular}[c]{@{}c@{}}winter peak load\\ - 2030 (Forecast)\end{tabular} \\ \hline
NYS & 32604 & 33978 & 25203 & 32495 \\ \hline
\end{tabular}
\label{CLCPApeakLoad}
\end{table}
\subsection{Nuclear power} \label{NuclearData}
The developed power grid contains the $4$ nuclear power facilities that are now functioning in NYS. They include Nine Mile I (629 MW), Nine Mile II (1299 MW), Ginna (581.7 MW), and FitzPatrick (854.5 MW), where Ginna power enters Zone B, and others enter Zone C. According to \cite{CARIS2019}, under the state support provided by Zero Emission Credit Requirements contained in the Clean Energy Standard, we expect these four nuclear facilities to continue operation until around 2030. We also assume that each nuclear power plant operates at 100 percent capacity for all hours of the modeling period. 

\subsection{Thermal power} \label{ThermalData}
We incorporate 227 fossil fuel generators into our model based on the NYS electric power grid model developed by \cite{liu2022open}. These are the main operational thermal generators in NYS in the year $2019$. In our base scenario, we assume all thermal generators are operational until the year $2030$, which is our study year. Later, for sensitivity analysis purposes, we retire some of them and investigate impacts on the operation of the power system and its risks. The required generator parameters are collected from the regional greenhouse gas initiative (RGGI) \cite{RGGIwebsite} as described in \cite{liu2022open}. The costs of power generation are also determined according to the approach employed by \cite{liu2022open}. We assume generator parameters are fixed over time. 


\subsection{Transmission lines} \label{TransmissionLineData} 
The NYS power grid is made up of 94 transmission lines; 68 lines connect the state's 11 zones and 8 lines connect NYISO to neighboring ISOs including IESO, ISONE, and PJM. We model the quantity of power that could be transferred across the power system according to NYISO’s Reliability Needs Assessment report \cite{RNA2020}. This report suggests  interface flow capacities for $2030$ based on the types of in-service cables, active reactors, expected transmission projects, generator retirement plans, and updated imports/exports plans. These limits are summarized in Table \ref{interfaceFlowLimits}, with the first 12 limits for internal zonal interface flows and the remainder for external trades.
\begin{table}[!t]
\centering
\caption{Interface flow limits (MW) for the year 2030 \cite{RNA2020}}
\begin{tabular}{|l|lll|l|lll|}
\hline
& Interface & Lower bound & Upper bound & & Interface & Lower bound & Upper bound \\ \hline
1 & A-B        	& -2200     		& 2200  & 10  & I-J        	& -4350     		& 4350 \\
2 & B-C        	& -1600     		& 1500  & 11  & I-K        	& -515    			& 1293 \\
3 & C-E        	& -5650       		& 5650  & 12  & E-FG       	& -3400     		& 5650 \\
4 & D-E        	& -1600     		& 2650  & 13  & A-IESO 		& -1700    			& 1300 \\
5 & E-F        	& -3925     		& 3925  & 14  & D-IESO 		& -300    			& 300  \\
6 & E-G        	& -1600    			& 2300  & 15  & F-ISONE		& -800    			& 800  \\
7 & F-G        	& -5400    			& 5400  & 16  & G-ISONE		& -600    			& 800  \\
8 & G-H        	& -7375     		& 7375  & 17  & C-PJM  		& -900    			& 500  \\
9 & H-I        	& -8450     		& 8450  & 18  & G-PJM  		& -1000    			& 1000 \\ 
\hline
\end{tabular}
\label{interfaceFlowLimits}
\end{table}
\section{Results} \label{Results}
We first focus on the spatiotemporal patterns of generated and dispatched power across the NYS system over the 40-year period of record (Section \ref{gridAnalysisResults}). This is followed by an analysis of power curtailment and congestion (Section \ref{CurtailmentResults}) and consideration of electricity price variability (Section \ref{LMPAnalysisResults}).  
\subsection{Dispatched Power, Generation, and Load Variability} \label{gridAnalysisResults}
\begin{figure}[t]
\includegraphics[width=\textwidth]{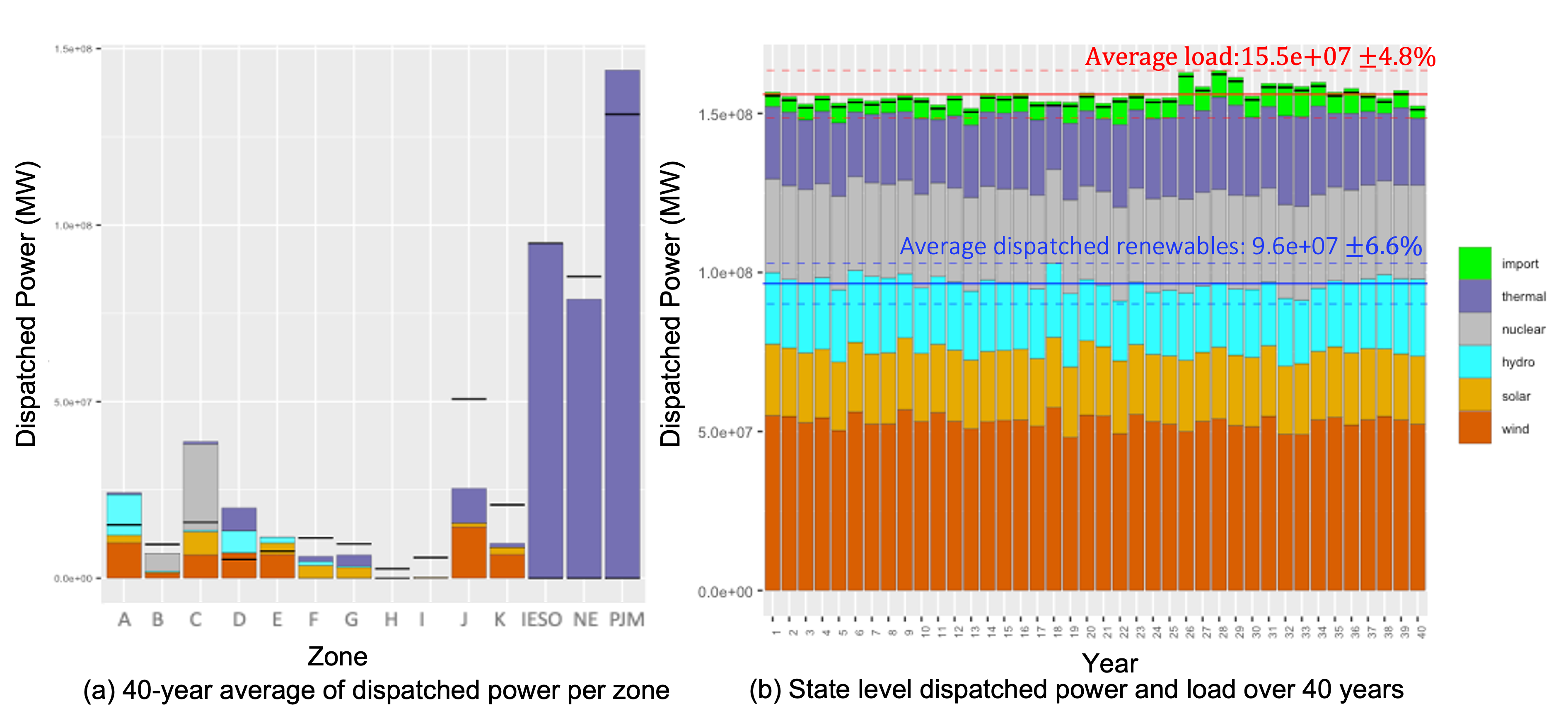}
\centering
\caption{Dispatched Power Comparison: Figure (a) shows the spatial patterns of average dispatched power across NYS over the 40-year period of record. Figure (b) represents NYS's annual dispatched power and electricity load over the $40$ simulated years. The black step-wise line illustrates the total load per year, and the region above this line indicates the lost power due to the charge and discharge of batteries. The green portion of each bar (labeled as an import) represents the annual imported power minus the exported power.}
\label{yearlyPowerGen1}
\end{figure}

Our analysis shows that on a long-term average and state-wide basis, dispatched power from renewable, nuclear, thermal, and imported sources account for $62\%$, $19\%$, $15\%$, and $4\%$ of the state load, respectively. However, these sources of dispatched power are not distributed evenly across the state. For renewable sources, on-shore wind power is concentrated upstate in Zones A, C, D, and E, offshore wind is concentrated off the coast of the New York City metropolitan area in Zones J and K, and dispatched hydropower is predominantly generated upstate in Zones A and D (Figure \ref{yearlyPowerGen1}-a). Dispatched solar power is more evenly distributed across zones, while nuclear power is concentrated mainly upstate in zones B and C. The consequence of this uneven distribution of generation is substantial reliance  on power transfer across zones. For example, upstate Zones A, C, D, and E dispatch more power than required to meet their own load in order to address power deficiencies downstate in Zones F, G, H, I, J, and K. Imported power also plays an important role in meeting demand in high load zones downstate and primarily comes from PJM to the southwest. 

The proportion of dispatched power used to meet load across the state also varies substantially over various timescales due to variability in climate, its impact on RES generation and energy demand, and their interaction with constraints in the power system. At an annual and state-wide scale, these interacting factors cause the dispatched renewables, including wind, solar, and hydropower, to range  +/- $6.6\%$ of the long-term average over the 40-year record (Figure \ref{yearlyPowerGen1}-b). For load, this range is +/-$4.8\%$. This degree of variability increases at shorter timescales and for different power sources. For instance, while the range of dispatched hydropower varies between +/- $13.6\%$ of its long-term average at an annual scale, this range increases to +/- $44\%$ at a monthly scale and +/- $89\%$ at a daily scale (Table \ref{Powerdev_dispatched}; also see Figure \ref{busLevelRES}). Dispatched wind and solar power vary considerably at shorter timescales, with a range reaching +/- $469\%$ for solar power and +/- $194\%$ for wind power at an hourly scale. (Percentage greater than $100\%$ represents the upper bound of the range.) Overall, dispatched hydropower is the most stable source at hourly, daily, and monthly scales, but is the most variable source at annual scales. Conversely, dispatched solar power is the most variable at hourly timescales but is the most stable renewable power source at an annual scale. This is because, in NYS, hydropower is primarily from great lakes, a big water source that does not change substantially in shorter time scales. Variability in the sum of dispatched power across all three renewable sources is lower across all timescales compared to any particular source, but still remains substantial across timescales.    
\begin{table}[!b]
\centering
\caption{Dispatched power average (MW) and deviation from average in various temporal resolutions}
\label{Powerdev_dispatched}
\begin{tabular}{|l|llll|}
\hline
                   & Yearly         	& Monthly       	& Daily          	& Hourly         	\\ \hline
Wind    & 5.3e+07 ±9.3\% 	& 4.4e+06 ±85\% 	& 1.5e+05 ±192\% 	& 6.1e+03 ±194\% 	\\
Solar   & 2.2e+07 ±6.6\% 	& 1.8e+06 ±63\% 	& 0.6e+05 ±107\% 	& 2.5e+03 ±469\% 	\\
hydro   & 2.1e+07 ±13.6\% 	& 1.8e+06 ±44\% 	& 0.6e+05 ±89\% 	& 1.8e+03 ±89\% 	\\
wind-solar-hydro   & 9.6e+07 ±6.6\% 	& 8.0e+06 ±38\% 	& 2.6e+05 ±94\% 	& 11.0e+03 ±135\% 	\\
thermal & 2.4e+07 ±23.6\% 	& 2.0e+06 ±136\% 	& 0.7e+05 ±319\% 	& 2.7e+03 ±583\% 	\\
Battery charge     & 0.6e+07 ±4.0\% 	& 0.5e+06 ±23\% 	& 0.18e+05 ±120\% 	& 0.7e+03 ±307\% 	\\
Battery discharge  & 0.5e+07 ±4.0\% 	& 0.4e+06 ±19\% 	& 0.14e+05 ±127\% 	& 0.6e+03 ±400\% 	\\
Import             & 1.2e+07 ±23\% 		& 1.0e+06 ±84\% 	& 0.3e+05 ±100\% 	& 2.4e+03 ±100\% 	\\
Export             & 0.6e+07 ±40\% 		& 0.5e+06 ±146\% 	& 0.2e+05 ±258\% 	& 1.6e+03 ±100\% 	\\
Load               & 15.5e+07 ±4.8\% 	& 12.9e+06 ±23\% 	& 4.2e+05 ±47\% 	& 17.7e+03 ±69\% 	\\ \hline
\end{tabular}
\end{table}

\begin{figure}[t]
\includegraphics[width=\textwidth]{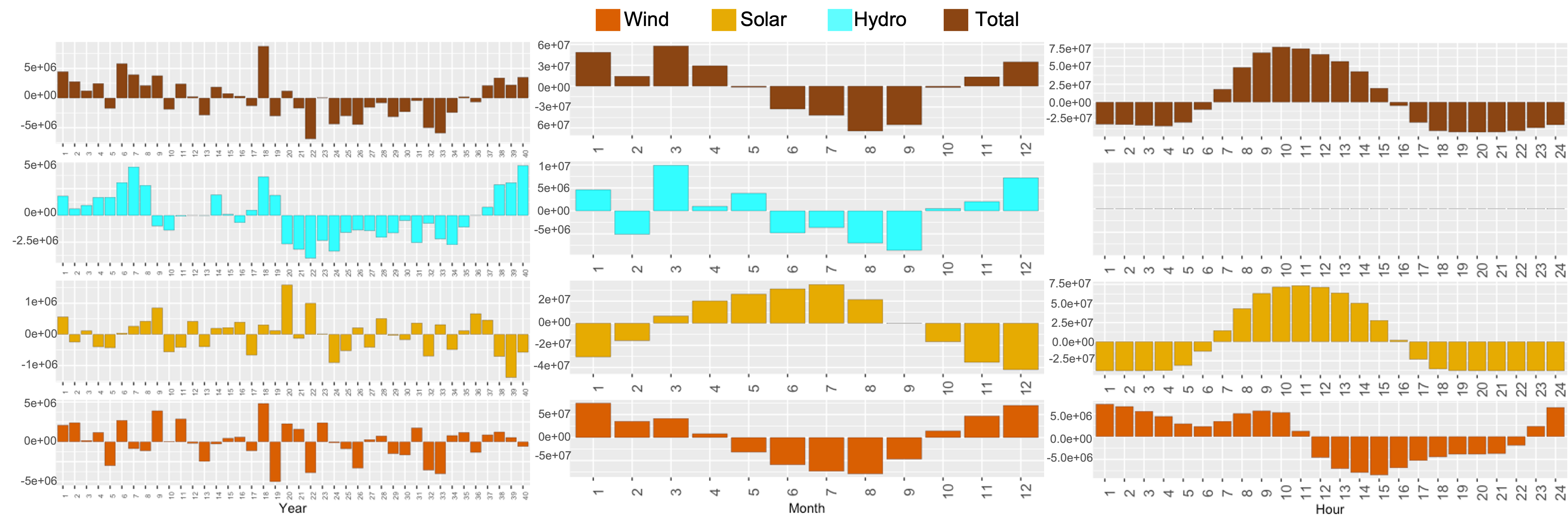}
\centering
\caption{The average yearly, monthly, and hourly deviation of generated renewables (MWh) from their long-term average}
\label{AvailableRES1}
\end{figure}

Some (though not all) of the variability in renewable dispatched power relates to variability in generation linked to weather, and this variability can exhibit different patterns across timescales depending on power source. This is visualized in Figure \ref{AvailableRES1}, which presents state-wide renewable generation over the 40-year record rather than dispatched power. Hydropower exhibits the most persistence in generation at an annual scale, with periods of above-average generation between years $1-20$ and again during years $35-40$, and periods of below-average generation between years $20-35$. This persistence is linked to slowly varying levels in the Great Lakes that drive generation at the Moses Niagara and Moses Saunders hydropower plants. 

Wind and solar generation do not exhibit similar persistence at the annual scale, with anomalies from the long-term average occurring randomly from year to year. However, these two sources do exhibit more persistent seasonal behavior in their generation, with wind generation reaching its maximum (minimum) in the cold (warm) season and solar generation exhibiting the opposite pattern. At hourly scales, solar generation exhibits the expected diurnal cycle, while wind power shows greater generation in the first half of the day and lower values in the second half. Hydropower exhibits no hourly variability in generation by design (see Section \ref{HydroData}).  

\subsection{Renewable Power Curtailment and Congestion}\label{CurtailmentResults}
\begin{figure}[t]
\includegraphics[width=\textwidth]{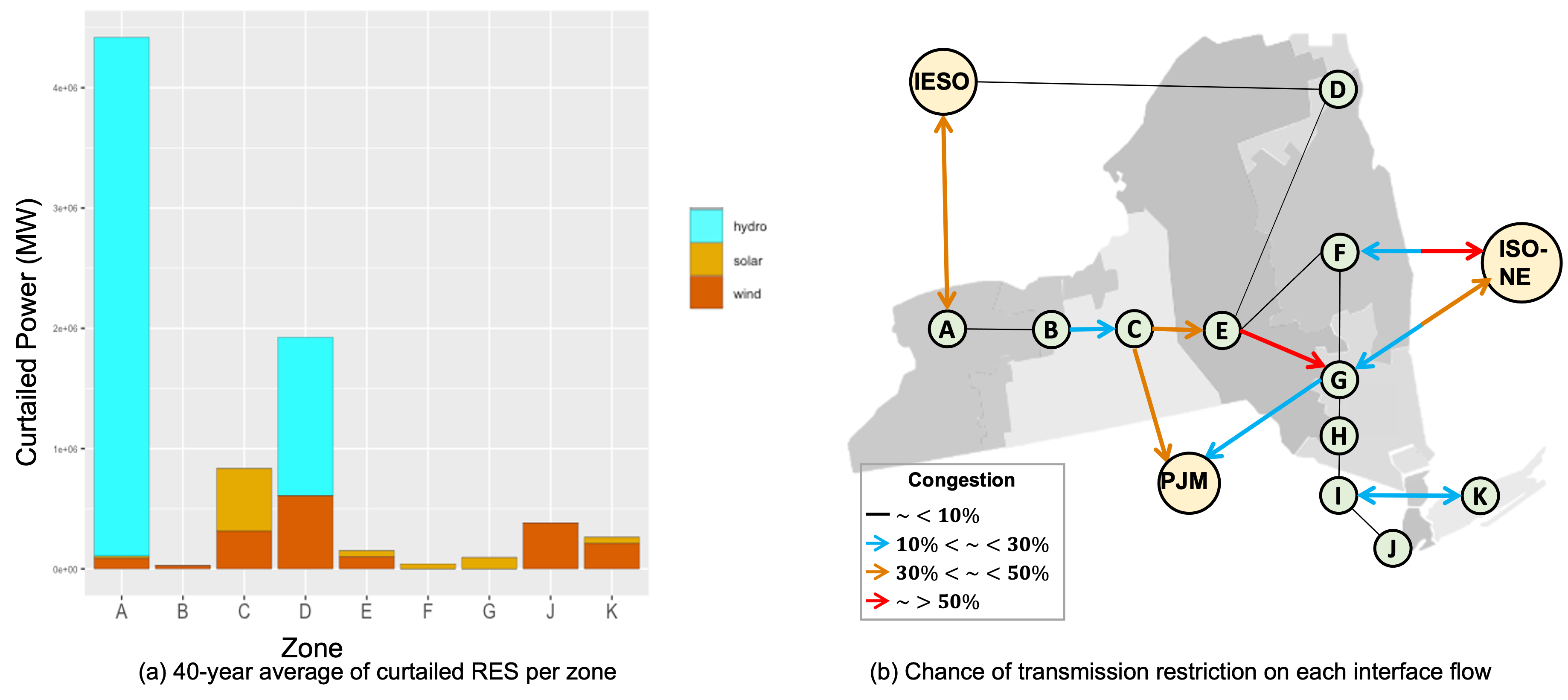}
\centering
\caption{Curtailed power and transmission congestion analysis}
\label{zonalCurtailmentCongestion}
\end{figure}

While some of the variability in dispatched renewable power is due to weather-driven generation patterns, a substantial proportion results from curtailment caused by system constraints. A large fraction of generated renewable power is curtailed in the system, with some renewable power unused in $62\%$ of hours in the 40-year record. However, curtailment varies widely by source and location (Figure \ref{zonalCurtailmentCongestion}-a). The most substantial curtailment occurs for hydropower in Zones A and D, accounting for $27\%$ and $17\%$ of total generation. This compares to just $1.8\%$ of total wind power being curtailed  and $2.2\%$ of solar power. 

Two factors primarily drive higher hydropower curtailments. First, hydropower is assumed to have slightly greater cost compared to wind and solar power (see Section \ref{Data}), so hydropower may be curtailed when wind and solar generation are sufficient to meet demand (see Figure \ref{dispatchedVScongestion}). Second, hydropower generation is concentrated in Zones A and D, and so is vulnerable to curtailment when congestion limits the transfer of power out of these zones. Interface flow limits account for $90\%$ of all renewable curtailment cases, while the remaining curtailment events are due to ramping constraints, imports of cheaper power (i.e., with negative LMP) from neighboring ISOs, and system-wide oversupply of renewable power. 

We explore the issue further in Figure \ref{zonalCurtailmentCongestion}-b, which shows congestion in various parts of the transmission system expressed as the fraction of time that flow reaches its limit at each interface. System congestion is mainly observed in three cases: (i) B$\rightarrow$C$\rightarrow$E$\rightarrow$G, (ii) K$\rightarrow$I, and (iii) at interfaces between NYISO and neighboring ISOs. The first case is due to high supply in zones A through E and high demand in zones F to K. This situation commonly arises when the combined dispatch of wind and solar power downstate is low and dispatched hydropower in Zones A and D is high, causing congestion and limiting transfer from upstate to downstate (see Figure \ref{dispatchedVScongestion}). The second case arises because of zone K's large offshore wind generation and its limited transmission capacity to other zones.
We note that curtailment in renewable generation is considerable despite the substantial capacity of batteries throughout the system (see Section \ref{StorageData}). Batteries supply an average of $3.4\%$ of the annual load and cut renewable energy curtailment by approximately $3\%$. However, the charging and discharging of batteries cause average losses of $0.6\%$ of the total annual load. Batteries have a greater influence at higher temporal frequencies; at the hourly time scale, batteries provide up to $27\%$\ of load, and during peak load hours, they deliver $10\%$ of load on average. Batteries in zones H, I, J, and D are cycled fully in more than $25\%$ of hours in the 40-year record, implying that adding additional storage capacity beyond the CLCPA scenario employed here would be beneficial (Figure \ref{batteryUsage}).  

\subsection{Electricity price evaluation} \label{LMPAnalysisResults}
\begin{figure}[!t]
\includegraphics[width=\textwidth]{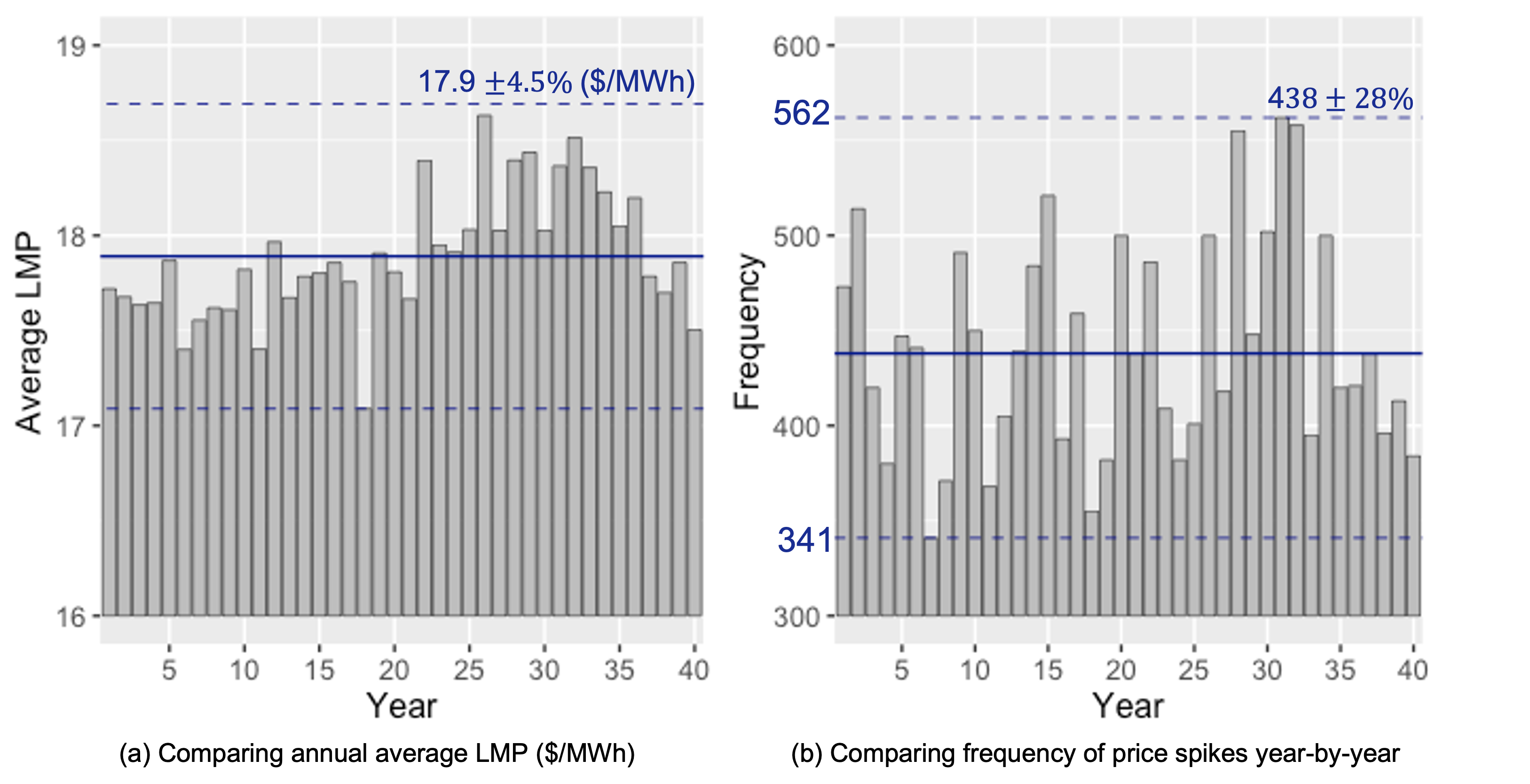}
\centering
\caption{Yearly average LMP and price spike frequency comparison. In each figure, the blue line shows the 40-year average, and the dashed lines are the upper/lower bounds of yearly values.}
\label{LMPandSpikeFreqComparison}
\end{figure}

\begin{figure}[t]
\includegraphics[width=\textwidth]{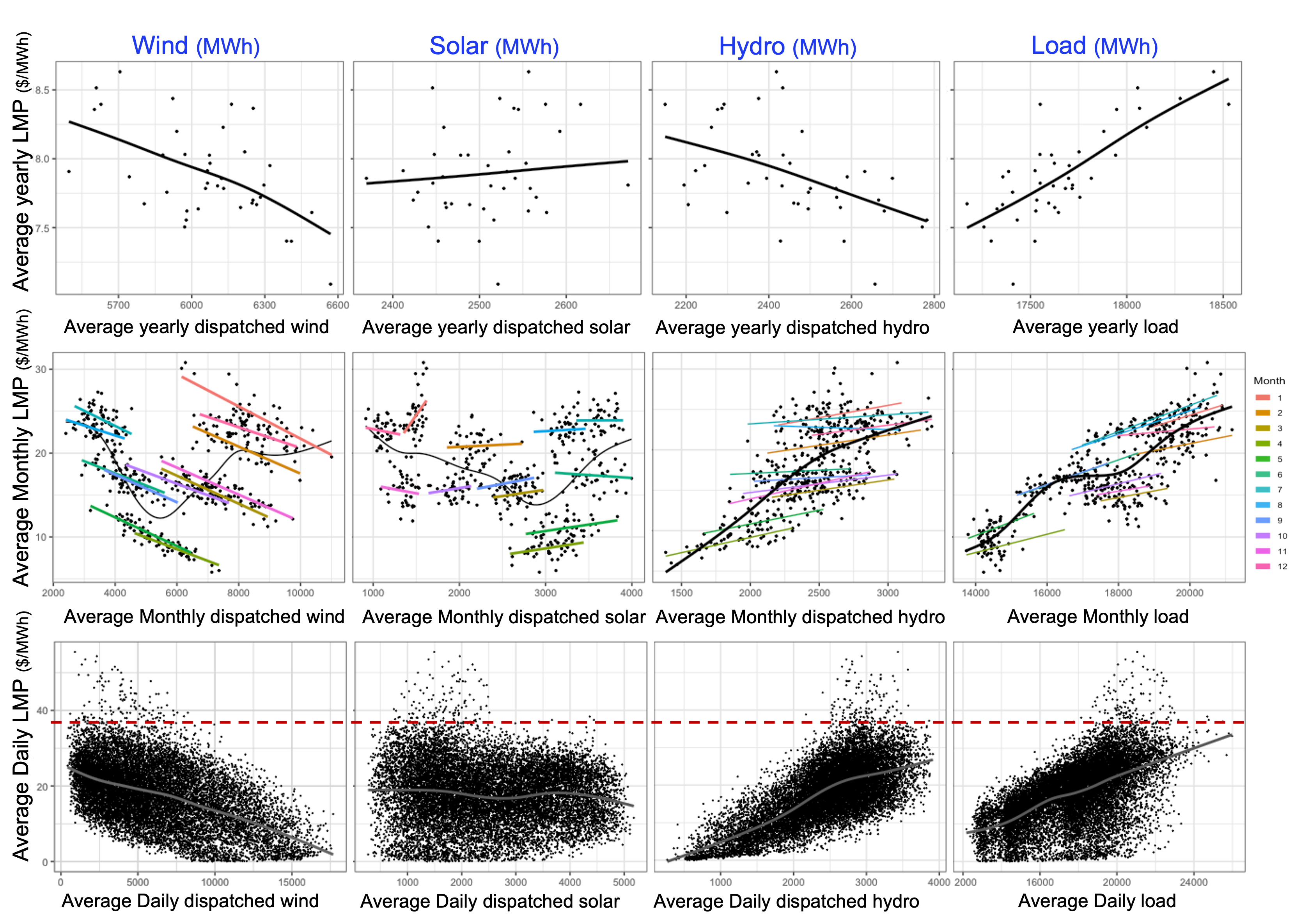}
\centering
\caption{Comparing the relationship between dispatched renewable energy and load and locational marginal prices (LMP) at various timescales. The solid black lines are the trend lines fitted on each scatterplot, and the red dashed line in the figures of daily data shows the price spike threshold. The solid colored lines in the monthly data figures show the trend within each month.}
\label{highLMPdrivers}
\end{figure}

The variability in dispatched power from renewable sources with low (or zero) marginal cost, in combination with variations in load, require more expensive thermal generators to occasionally mitigate supply-demand imbalances in the system. This dynamic leads to variability in LMPs across the system, which we investigate here. At a state-wide and annual average scale, LMPs vary by +/-$4.5\%$ of their long-term average across the 40-year period of record (Figure \ref{LMPandSpikeFreqComparison}-a). We note a period in years 20-40 that exhibit higher annual average LMPs corresponding to a period with concurrent negative anomalies in multiple renewable sources, especially annual hydropower and wind power (see Figure \ref{AvailableRES1}). However, variability in LMPs is  greater when viewed at finer temporal scales. For instance, we define a \textit{price spike} as an hourly price being above $\$33.5$/MWh, which is the $95$th percentile of all hourly LMPs during the 40-year study period. We find that the annual frequency of price spikes varies by +/-$28\%$ of the long-term average across the 40-year record (Figure \ref{LMPandSpikeFreqComparison}-b). Interestingly, although annual average LMPs and annual price spike frequencies often occur in similar years, there is only a moderate correlation between these two variables. This suggests that the drivers leading to variability in these two price statistics differ across years.  
To better understand this result, we investigate the link between average LMPs at different timescales and the factors influencing price spikes, including dispatched renewable energy and load (Figure \ref{highLMPdrivers}). Unsurprisingly, at all timescales, an increase in load leads to an increase in price. However, we find that the variability of prices at monthly and annual timescales is mainly driven by variability in wind and hydropower generation. Conversely, price spikes at daily (and hourly) timescales clearly are driven by multiple factors occurring simultaneously, including low dispatched wind and solar power and high levels of dispatched hydropower during periods of high load.

\begin{figure}[t]
\includegraphics[width=\textwidth]{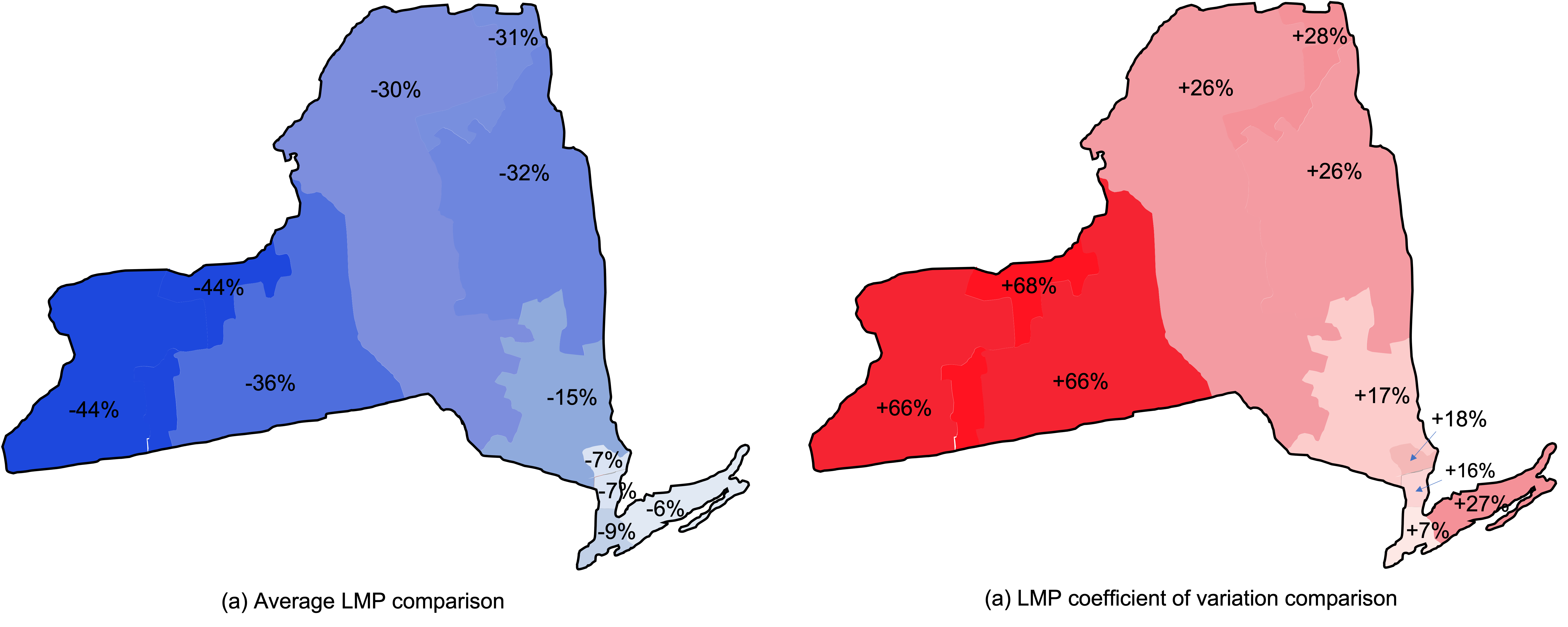}
\centering
\caption{LMP average and variability comparison between the NY grid with large-scale renewable integration and the simulated grid in 2019: Figure (a) shows the decrease in average LMP due to the considerable addition of renewable resources across zones, and Figure (b) shows the increase in the coefficient of variation of LMPs across zones.}
\label{zonalLMPComparison} 
\end{figure}

Finally, for context, we compare the variability of state-level weighted LMPs from our modeled results with the variability of LMPs in the simulated model for 2019, when renewable sources constitute a far smaller share of total load (see Figure \ref{zonalLMPComparison}). Our results show that large-scale integration of renewable sources into the grid decreases zonal LMPs, particularly in upstate zones (i.e., A-F) where there are more RES. However, price variability increases, with the coefficient of variation of state-level LMPs increasing between $12\%$ to $28\%$, with an average of $17\%$. This is especially true in zones with large amounts of renewable energy generation.
\section{Discussion} \label{Discussion}
This paper explores the potential outcomes associated with the increased build-out of renewable energy sources, such as those proposed in many countries and regions. While the approach is generalizable, our analysis uses the case of the NYS CLCPA to illustrate and explore performance over a range of spatial and temporal scales. Results of our analysis show that the spatial allocation of renewable resources will necessitate a considerable increase in the need for power transfer across zones. The curtailment of renewable power in about two-thirds of all hours in the 40-year simulated data shows that the resource availability does not align with load patterns, and the system needs substantially more storage capacity to make efficient use of the planned renewable resource.  

In addition to spatial variability, the power dispatch varies substantially over time due to variability in climate, driving RES generation and electricity demand, and the interaction of generation and demand with constraints in the power system. Different renewable power resources exhibit distinct temporal variability patterns; hydropower is the most stable energy source at shorter time scales and solar and wind power at longer time scales. Understanding the variability and uncertainty at a range of time scales is necessary to ensure the right mix of resources to mitigate potential issues of power system reliability.

Perhaps unsurprisingly, the annual frequency of electricity price spikes depends on variability in renewable energy resources and load and their interactions with power system infrastructure limits, specifically transmission constraints. However, the drivers of increased volatility vary across different time scales. Also, the planned renewable resource mix causes heterogeneity in price changes across the space. Spatial electricity price heterogeneity can affect energy consumption patterns, the adoption of renewable energy sources 
, and the electricity grid's stability, and create or exacerbate disparities in access to affordable energy. Therefore, the spatial patterns of impacts on electricity prices should be carefully considered by policymakers and energy regulators when designing renewable expansion policies. Increasing the grid's transmission capacity, promoting further energy storage, implementing demand response programs, and adopting more renewable energy sources, particularly in regions with higher electricity prices, are potential levers to reduce price heterogeneity through evenly balancing demand and supply across different regions, though the effectiveness of these measures depends on whether their implementation is responsive to the underlying spatial drivers of price volatility.  

As we have seen, the impact of variability in RES on  system reliability and electricity prices is sensitive to the structure of the power system. Specific outcomes, therefore, depend on differences between grid topologies, but are also influenced by model assumptions underlying the analysis. We analyzed the sensitivity of our findings to the underlying assumptions. Running the model for one sample year (i.e., year 32 with very large LMPs) demonstrates that the results are robust to our assumptions about the capacity and locations of wind, solar, and battery facilities. Under twelve alternative scenarios for this specific year, average LMP, price spike frequency, and total renewable generation, respectively, deviate up to $0.5\%$, $3\%$, and $1\%$ from the base case scenario. The results, however, show higher sensitivity to assumptions made regarding the renewable capacity allocations across zones taken from \cite{CARIS2019}, and also state-level capacity targets codified in the CLCPA. The reader is directed to the Appendix for more detail on the sensitivity analysis. 

While the results of this study highlight essential considerations and previously unreported risks to large-scale integration of renewable energy with a specific case study, this work has several caveats. First, even though we adopt an abstracted and  validated representation of the power system, the model uses a relatively simplistic representation of market mechanisms, as well as excluding power losses due to transfers, dynamic transmission ratings, variability in solar panel efficiencies, and the impacts of renewable resource forecast uncertainty on battery operations, among other factors. Additionally, most of the assumptions made in our modeling framework were designed to be optimistic and are based on the assumption that the CLCPA targets would be reached. However, these goals are subject to critical challenges facing large-scale and rapid development and integration of renewable generation. The accuracy of the analysis could potentially be improved with the refinement of the developed load prediction model. The model can be further improved by (1) formulating auto-correlation between adjacent hourly loads and (2) adding the hour of the day as another feature vector. This may ignore the auto-correlation effect, but it may also avoid error propagation. Finally, we model power generation and price in neighboring ISOs based on their operation in 2019. Presumably, changes occurring in neighboring grids, including the addition of more renewable energy and the retirement of thermal generators, will affect the dynamics of transfers over these interfaces. Thus, the power flow and generation scheduling represented in this analysis may not capture the complex interactions of this evolution. 


\section{Summary and Conclusion} \label{Conclusion}
This study characterized the spatiotemporal variability in renewable energy sources, load, and the resulting electricity price volatility based on a realistic representation of the NYS power grid while integrating additional wind and solar power plants proposed for NYS through 2030 by the CLCPA. Using the historical record of meteorological variables over a 40-year time window, renewable generation and the electricity demand are simulated with the hourly resolution to maintain the co-variability between these inputs. The proposed modeling framework facilitates analysis of system operational performance under a wide range of weather conditions and with a comprehensive representation of renewable energy variability and electricity price volatility.

Our results demonstrate that each renewable energy source can deviate up to $17\%$ above and below average annually, while hybridization of wind, solar, and hydropower reduce the net variation to $8\%$. On an hourly basis, renewable volatility is substantially greater, and it may deviate $100\%$ above and below average, creating challenges for infrastructure and operational reliability. The effects of renewable energy on prices are driven by different factors across time scales. For example, in this case study, renewable variability induces a $9\%$ variation in annual average electricity prices and a $56\%$ variation in the frequency of price spikes. While yearly average price volatility is mostly caused by differences in hydropower availability, daily and hourly price volatility is caused by fluctuations in solar and wind availability.

Understanding the RES co-variability and its consequent risks at various temporal and spatial resolutions informs development and investment in power system infrastructure. It helps power system operators develop better regulation and dispatch policies to be well-prepared for changes in power production posed by extensive RES deployment. The results of this study provide additional insight into the necessity for further energy storage and transmission capacity. The proposed framework can be expanded to different years, for example $2040$, to study the risks of $100\%$ renewable scenario under potential shifts in the climate over time. Our framework can be extended to other power systems, particularly those with substantial contributions from hydroelectric generation, such as the eastern US.

\section*{Acknowledgment}
The authors gratefully acknowledge financial support from the U.S. Department of Agriculture (USDA) under Grant Number 2019-67019-30122. Any opinions in the paper are those of the authors and do not necessarily reflect the opinion of the sponsor. We also thank Kenji Doering for providing us with Moses Niagara hydropower data and help with developing models for renewable power generation and Kyla Semmendinger for providing us with Moses Saunders hydropower data.

\bibliographystyle{elsarticle-num}
\bibliography{refrences.bib}

\begin{thebibliography}{10}
\expandafter\ifx\csname url\endcsname\relax
  \def\url#1{\texttt{#1}}\fi
\expandafter\ifx\csname urlprefix\endcsname\relax\def\urlprefix{URL }\fi
\expandafter\ifx\csname href\endcsname\relax
  \def\href#1#2{#2} \def\path#1{#1}\fi

\bibitem{masson2022global}
V.~Masson-Delmotte, P.~Zhai, H.-O. P{\"o}rtner, D.~Roberts, J.~Skea, P.~R.
  Shukla, et~al., Global Warming of 1.5° C: IPCC Special Report on Impacts of
  Global Warming of 1.5° C above Pre-industrial Levels in Context of
  Strengthening Response to Climate Change, Sustainable Development, and
  Efforts to Eradicate Poverty, Cambridge University Press, 2022.

\bibitem{IEA2022CO2emissions}
{IEA} {G}lobal {E}nergy {R}eview: {CO2} {E}missions in 2021, Tech. rep.,
  International Energy Agency (IEA), {Paris}, License: CC BY 4.0 (2022).

\bibitem{USNetZeroPlan2021}
{The Long-term Strategy of the United States: Pathways to Net-Zero Greenhouse
  Gas Emissions by 2050}, Tech. rep., The United States Department of State and
  the United States Executive Office of the President, Washington DC. (2021).

\bibitem{alshawaf2020solar}
M.~Alshawaf, R.~Poudineh, N.~S. Alhajeri, Solar {PV} in {Kuwait}: The effect of
  ambient temperature and sandstorms on output variability and uncertainty,
  Renewable and Sustainable Energy Reviews 134 (2020) 110346.

\bibitem{ela2013impacts}
E.~Ela, V.~Diakov, E.~Ibanez, M.~Heaney, {Impacts of Variability and
  Uncertainty in Solar Photovoltaic Generation at Multiple Timescales}, Tech.
  rep., National Renewable Energy Lab. (NREL), Golden, CO (United States)
  (2013).

\bibitem{naegele2020climatology}
S.~Naegele, T.~McCandless, S.~Greybush, G.~Young, S.~Haupt, M.~Al-Rasheedi,
  Climatology of wind variability for the {Shagaya} region in {Kuwait},
  Renewable and Sustainable Energy Reviews 133 (2020) 110089.

\bibitem{abunima2022two}
H.~Abunima, W.-H. Park, M.~B. Glick, Y.-S. Kim, Two-stage stochastic
  optimization for operating a renewable-based microgrid, Applied Energy 325
  (2022) 119848.

\bibitem{quan2019survey}
H.~Quan, A.~Khosravi, D.~Yang, D.~Srinivasan, A survey of computational
  intelligence techniques for wind power uncertainty quantification in smart
  grids, IEEE Transactions on Neural Networks and Learning Systems 31~(11)
  (2019) 4582--4599.

\bibitem{pereira2019effect}
P.~Pereira~da Silva, P.~Horta, The effect of variable renewable energy sources
  on electricity price volatility: the case of the {Iberian} market,
  International Journal of Sustainable Energy 38~(8) (2019) 794--813.

\bibitem{bett2016climatological}
P.~E. Bett, H.~E. Thornton, The climatological relationships between wind and
  solar energy supply in {Britain}, Renewable Energy 87 (2016) 96--110.

\bibitem{monforti2014assessing}
F.~Monforti, T.~Huld, K.~B{\'o}dis, L.~Vitali, M.~D'isidoro,
  R.~Lacal-Ar{\'a}ntegui, Assessing complementarity of wind and solar resources
  for energy production in {Italy}. a monte carlo approach, Renewable Energy 63
  (2014) 576--586.

\bibitem{mohammadi2018study}
K.~Mohammadi, N.~Goudarzi, Study of inter-correlations of solar radiation, wind
  speed and precipitation under the influence of {El Ni{\~n}o} {Southern}
  {Oscillation} {(ENSO)} in {California}, Renewable Energy 120 (2018) 190--200.

\bibitem{zhao2021new}
N.~Zhao, F.~You, {New York S}tate's 100\% renewable electricity transition
  planning under uncertainty using a data-driven multistage adaptive robust
  optimization approach with machine-learning, Advances in Applied Energy 2
  (2021) 100019.

\bibitem{huang2021economic}
K.~Huang, P.~Liu, B.~Ming, J.-S. Kim, Y.~Gong, Economic operation of a
  wind-solar-hydro complementary system considering risks of output shortage,
  power curtailment and spilled water, Applied Energy 290 (2021) 116805.

\bibitem{bird2016wind}
L.~Bird, D.~Lew, M.~Milligan, E.~M. Carlini, A.~Estanqueiro, D.~Flynn,
  E.~Gomez-Lazaro, H.~Holttinen, N.~Menemenlis, A.~Orths, et~al., Wind and
  solar energy curtailment: A review of international experience, Renewable and
  Sustainable Energy Reviews 65 (2016) 577--586.

\bibitem{ciarreta2020renewable}
A.~Ciarreta, C.~Pizarro-Irizar, A.~Zarraga, Renewable energy regulation and
  structural breaks: An empirical analysis of {Spanish} electricity price
  volatility, Energy Economics 88 (2020) 104749.

\bibitem{rintamaki2017does}
T.~Rintam{\"a}ki, A.~S. Siddiqui, A.~Salo, Does renewable energy generation
  decrease the volatility of electricity prices? an analysis of {Denmark} and
  {Germany}, Energy Economics 62 (2017) 270--282.

\bibitem{liu2022open}
M.~V. Liu, B.~Yuan, Z.~Wang, J.~A. Sward, K.~M. Zhang, C.~L. Anderson, An open
  source representation for the {NYS} electric grid to support power grid and
  market transition studies, IEEE Transactions on Power Systems (2022).

\bibitem{BillS6599}
{N}ew {Y}ork {S}tate {S}enate {B}ill {S6599},
  \url{https://www.nysenate.gov/legislation/bills/2019/s6599}, (accessed
  December 28, 2022).

\bibitem{CESA_cleanEnergyStates}
{C}lean {E}nergy {S}tates {A}lliance,
  \url{https://www.cesa.org/projects/100-clean-energy-collaborative/guide/table-of-100-clean-energy-states/},
  (accessed November 15, 2022).

\bibitem{CLCPA}
{C}limate {L}eadership and {C}ommunity {P}rotection {A}ct {(CLCPA)},
  \url{https://climate.ny.gov/-/media/Project/Climate/Files/Draft-Scoping-Plan.pdf},
  (accessed November 15, 2022).

\bibitem{NYISOphaseI2019}
{N}ew {Y}ork {ISO} {C}limate {C}hange {I}mpact {S}tudy - {Phase I}: {L}ong-term
  {L}oad {I}mpact, Tech. rep., Itron Incorporation (2019).

\bibitem{NYISOphaseII2020}
P.~J. Hibbard, C.~Wu, H.~Krovetz, T.~Farrell, J.~Landry, {Climate Change Impact
  Phase {II} - An Assessment of Climate Change Impacts on Power System
  Reliability in New York State}, Tech. rep., Analysis Group (2020).

\bibitem{GoldBook2021}
{NYISO Gold Book - Load and Capacity Data}, Tech. rep., New York Independent
  System Operator (2021).

\bibitem{CARIS2019}
{Congestion Assessment and Resource Integration Study}, Tech. rep., New York
  Independent System Operator (2019).

\bibitem{RNA2020}
{Reliability Needs Assessment}, Tech. rep., New York Independent System
  Operator (2020).

\bibitem{Hoen2018united}
B.~D. Hoen, J.~E. Diffendorfer, J.~T. Rand, L.~A. Kramer, C.~P. Garrity, H.~E.
  Hunt, {United States} wind turbine database v5.1 (july 29, 2022), Tech. rep.,
  {U.S.} Geological Survey, American Wind Energy Association, and Lawrence
  Berkeley National Laboratory data release (2018).

\bibitem{NYSERDA_RenewableProjects}
{L}arge-scale {R}enewable {P}rojects {R}eported by {NYSERDA}: {B}eginning 2004,
  \url{https://data.ny.gov/Energy-Environment/Large-scale-Renewable-Projects-Reported-by-NYSERDA/dprp-55ye/data},
  (accessed September 1, 2022).

\bibitem{NRELWTK}
{N}ational {R}enewable {E}nergy {L}ab ({NREL}) {W}ind {I}ntegration {N}ational
  {D}ataset (wind) {T}oolkit (wtk), \url{https://data.nrel.gov/submissions/54},
  (accessed Jun 28, 2021).

\bibitem{nyserda_offshoreWind}
{NY} {O}ff-shore {W}ind {P}rojects,
  \url{https://www.nyserda.ny.gov/All-Programs/Offshore-Wind/Focus-Areas/NY-Offshore-Wind-Projects},
  (accessed August 1, 2022).

\bibitem{doering2023evaluating}
K.~Doering, C.~L. Anderson, S.~Steinschneider, Evaluating the intensity,
  duration, and frequency of flexible energy resources needed in a
  zero-emission, hydropower reliant power system, Oxford Open Energy (2023)
  oiad003.

\bibitem{draxl2015wind}
C.~Draxl, A.~Clifton, B.-M. Hodge, J.~McCaa, The wind integration national
  dataset (wind) toolkit, Applied Energy 151 (2015) 355--366.

\bibitem{bloom2016eastern}
A.~Bloom, A.~Townsend, D.~Palchak, J.~Novacheck, J.~King, C.~Barrows,
  E.~Ibanez, M.~O'Connell, G.~Jordan, B.~Roberts, et~al., {Eastern Renewable
  Generation Integration Study}, Tech. rep., National Renewable Energy
  Lab.(NREL), Golden, CO ({United States}) (2016).

\bibitem{perpinan2007calculation}
O.~Perpi{\~n}an, E.~Lorenzo, M.~Castro, On the calculation of energy produced
  by a {PV} grid-connected system, Progress in Photovoltaics: Research and
  Applications 15~(3) (2007) 265--274.

\bibitem{meyer2017evaluating}
E.~S. Meyer, G.~W. Characklis, C.~Brown, Evaluating financial risk management
  strategies under climate change for hydropower producers on the {G}reat
  {L}akes, Water Resources Research 53~(3) (2017) 2114--2132.

\bibitem{semmendinger2022establishing}
K.~Semmendinger, D.~Lee, L.~Fry, S.~Steinschneider, Establishing opportunities
  and limitations of forecast use in the operational management of highly
  constrained multiobjective water systems, Journal of Water Resources Planning
  and Management 148~(8) (2022) 04022044.

\bibitem{NYISOfuelMix}
{N}ew {Y}ork {ISO} — {E}nergy and {M}arket {O}perational {D}ata,
  \url{https://www. nyiso.com/energy-market-operational-data}, (accessed Jun
  28, 2021).

\bibitem{GoldBook2019}
{NYISO Gold Book - Load and Capacity Data}, Tech. rep., New York Independent
  System Operator (2019).

\bibitem{NYISOload}
New york {ISO} - {L}oad {D}ata, \url{https://www.nyiso.com/load-data},
  (accessed Jun 28, 2021).

\bibitem{chow1991user}
J.~Chow, G.~Rogers, User manual for power system toolbox, Version 3 (1991)
  1991--2008.

\bibitem{RGGIwebsite}
{T}he {R}egional {G}reenhouse {G}as {I}nitiative ({RGGI}),
  \url{https://www.rggi.org/}, (accessed Jun 28, 2021).

\bibitem{Jeff2000}
{New York State Area for Consideration for the Potential Locating of Offshore
  Wind Energy Areas}, Tech. rep., the New York State Energy Research and
  Development Authority ({NYSERDA}), the New York Department of State ({DOS}),
  and the New York Department of Environmental Conservation ({DEC}) (2017).

\bibitem{NREL_solar}
{S}olar {P}ower {D}ata for {I}ntegration {S}tudies,
  \url{https://www.nrel.gov/grid/solar-power-data.html}, (accessed August 1,
  2022).

\end{thebibliography}

\newpage

\appendix
\section*{Appendix} \label{Appendix}
In this Appendix, we provide further detail about our data and results.  

\begin{table}[!b]
\centering
\footnotesize
\begin{tabular}{|ll|}
\hline
\textbf{Sets and Indexes}&  \\
$\mathcal{T}$ & set of time period, $t \in \mathcal{T}$ \\ 
$\mathcal{N}$ & set of all nodes/buses in the transmission system, $k \in \mathcal{N}$ \\ 
$\mathcal{G}(k)$ & set of generators connected to node $k$, $g \in \mathcal{G}(k)$ \\ 
$\mathcal{R}$ & set of renewable types including hydro, solar, and wind, $\mathcal{R}=\{H,S,W \}$ \\ 
$\mathcal{L}$ & set of power flow lines in the transmission system, $l=(m,k) \in \mathcal{L}$ \\ 
$\mathcal{I}$ & set of interfaces in the transmission system, $i \in \mathcal{I}$ \\ 
$\mathcal{I}_k$ & set of lines flow in to bus $k$\\
$\mathcal{O}_k$ & set of lines flow out of bus $k$\\
$\mathcal{IF}_i$ & set of lines in zonal interface $i$, $\mathcal{IF}_i \subset \mathcal{L}$\\
\textbf{Parameters} &  \\

$\overline{R}_{k,g} / \underline{R}_{k,g}$ & upper/lower ramp rate limit of transmission system generator $g$ \\
$\overline{P}_{k,g} / \underline{P}_{k,g}$ & generation upper/lower bound of transmission system generator $g$ \\
$\overline{L} / \underline{L}$ & upper/lower bound of transmission system line \\
$\overline{L}_{\mathcal{IF}_i} / \underline{L}_{\mathcal{IF}_i}$ & upper/lower bound of interface flow \\
$C_{k,g}(t) / C_{k,g}^{c}$ & linear/constant cost coefficient of generator $g$ at time $t$ \\
$E_{k}(t)$ & active demand for bus $k$ during time period $t$ \\
$SE_{k}$ & round-trip efficiency of the storage unit at bus $k$ \\
$S_{k}$ & maximum level of stored energy (in MWh) for the storage unit at bus $k$ \\
$P_{k}$ & power capacity (in MW) of the storage unit at bus $k$ \\
\textbf{Variables} &  \\
$p_{k,g}(t)$ & generation of generator $g$ during time $t$ \\
$e_{l}(t)$ & power flow of branch $l$ during time $t$ \\
$l_{k}(t)$ & unmet electricity demand at bus $k$ during time period $t$ \\
$\theta_{k}(t)$ & phase angle of bus $k$ during time $t$ \\
$ps_{k}(t)$ & power injections of the storage unit at bus $k$ during time $t$ \\
$psc_{k}(t), psd_{k}(t)$ & charge/discharge power injections of the storage unit at bus $k$ during time $t$ \\
$s_{k}(t)$ & amount of stored energy in the storage unit at bus $k$ at the beginning of hour $t$ \\
\hline
\end{tabular}
\end{table}

\subsection*{Data} \label{AppendixData}
\textbf{Wind Power}: To designate the potential locations of land-based wind turbines, we use the National Renewable Energy Lab (NREL) Wind Integration National Dataset (WIND) Toolkit (WTK) site dataset. This data represents the best potential locations for wind sites selected based on geography and wind resource data. Each site is a 2km*2km grid cell that, based on its buildable land area, contains one to eight 2-MW land-based wind turbines with a hub height of $100$ m. Depending on the annual average wind speed, each site contains a specific class of wind turbines. Accordingly, a separate power curve is assigned to each site class in order to convert wind speed into normalized power. In NYS, $2859$ potential wind power production sites with a total capacity of $37712$ MW are detected.

To select and add wind sites to our model, we first use the United States Wind Turbine Database \cite{Hoen2018united} from the United States Geological Survey (USGS). This dataset includes all land-based and offshore wind sites in the United States. The total capacity of all operational turbines in NYS is 1985 MW. We use another dataset, including the large-scale renewable projects beginning in 2004, reported by New York State Energy Research and Development Authority (NYSERDA) \cite{NYSERDA_RenewableProjects}. The dataset includes the ten large-scale land-based wind sites with a total capacity of $1392$ MW, which are under development and will be operational by $2025$. 

\renewcommand{\thefigure}{A\arabic{figure}}
\setcounter{figure}{0}

\begin{figure}[!t]
\includegraphics[width=\textwidth]{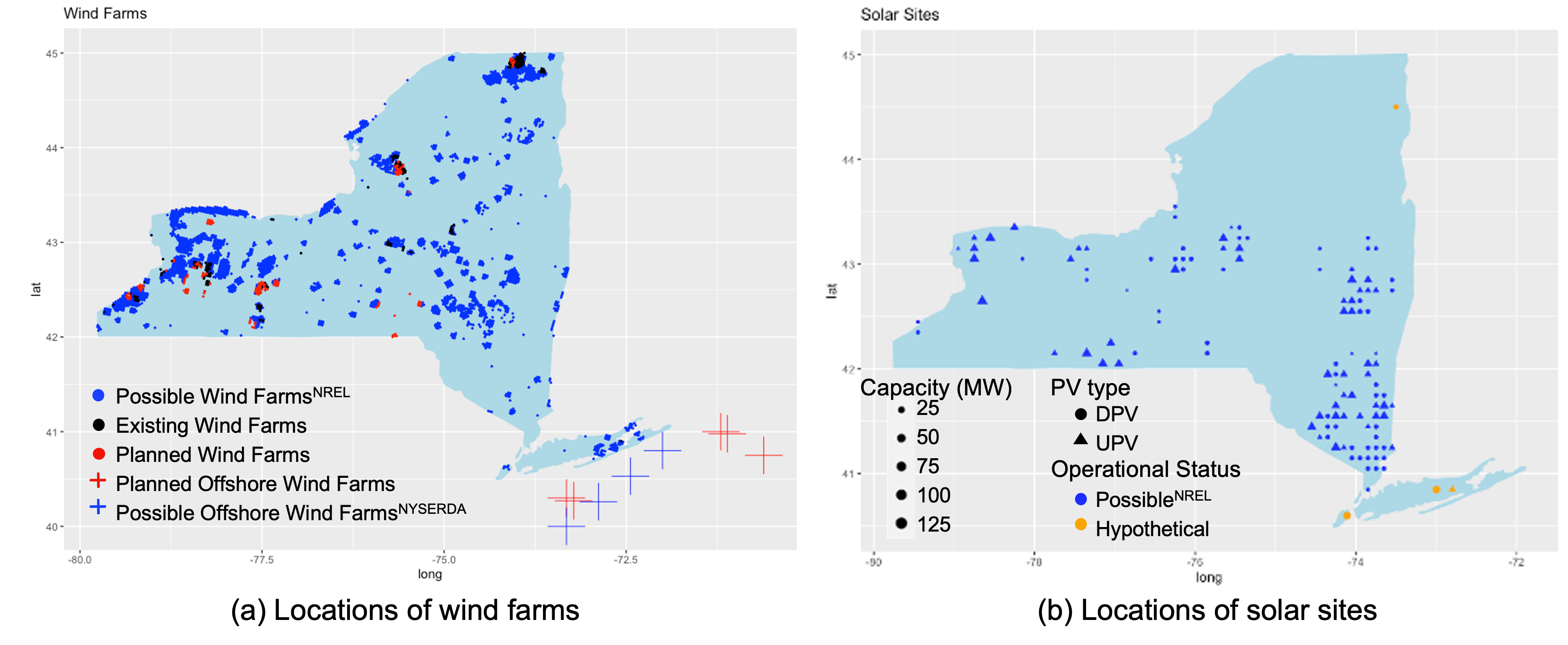}
\centering
\caption{Locations of all wind farms and solar sites in NYS}
\label{windSolarLocations}
\end{figure}

We pick offshore wind sites based on offshore wind projects reported by NYSERDA. To help meet New York’s nation-leading renewable energy targets, NYSERDA issues competitive solicitations for offshore wind energy and contracts with offshore wind developers to purchase offshore renewable energy certificates (ORECs). According to \cite{nyserda_offshoreWind}, the five active offshore wind projects off the coast of Long Island include South fork wind, Empire wind 1, Sunrise wind, Empire wind 2, and Beacon wind. In total, their capacity will be 4300 MW, and they will be entering commercial operation by 2028. In a joint work undertaken by NYSERDA, the New York Department of State (DOS), and the New York Department of Environmental Conservation (DEC), a massive area south of Long Island is studied to identify other zones for future offshore wind development. According to \cite{Jeff2000}, four zones are identified, each capable of supporting at least $800$ MW of future offshore wind development. We designate these areas as the potential locations for future offshore wind farms to supply the unplanned offshore wind power capacity that should be generated by $2030$. Figure \ref{windSolarLocations}-a represents the location of all land-based and offshore wind sites. We integrate all operational and under-development wind sites into our model. This includes $212$ operational and $148$ under-development wind farms with total capacities of $1985$ (MW) and  $1392$ (MW). We then select from among the potential land-based and offshore wind sites for extra MWs of wind power and to meet the CLCPA targets.  

\begin{figure}[t]
\includegraphics[width=\textwidth]{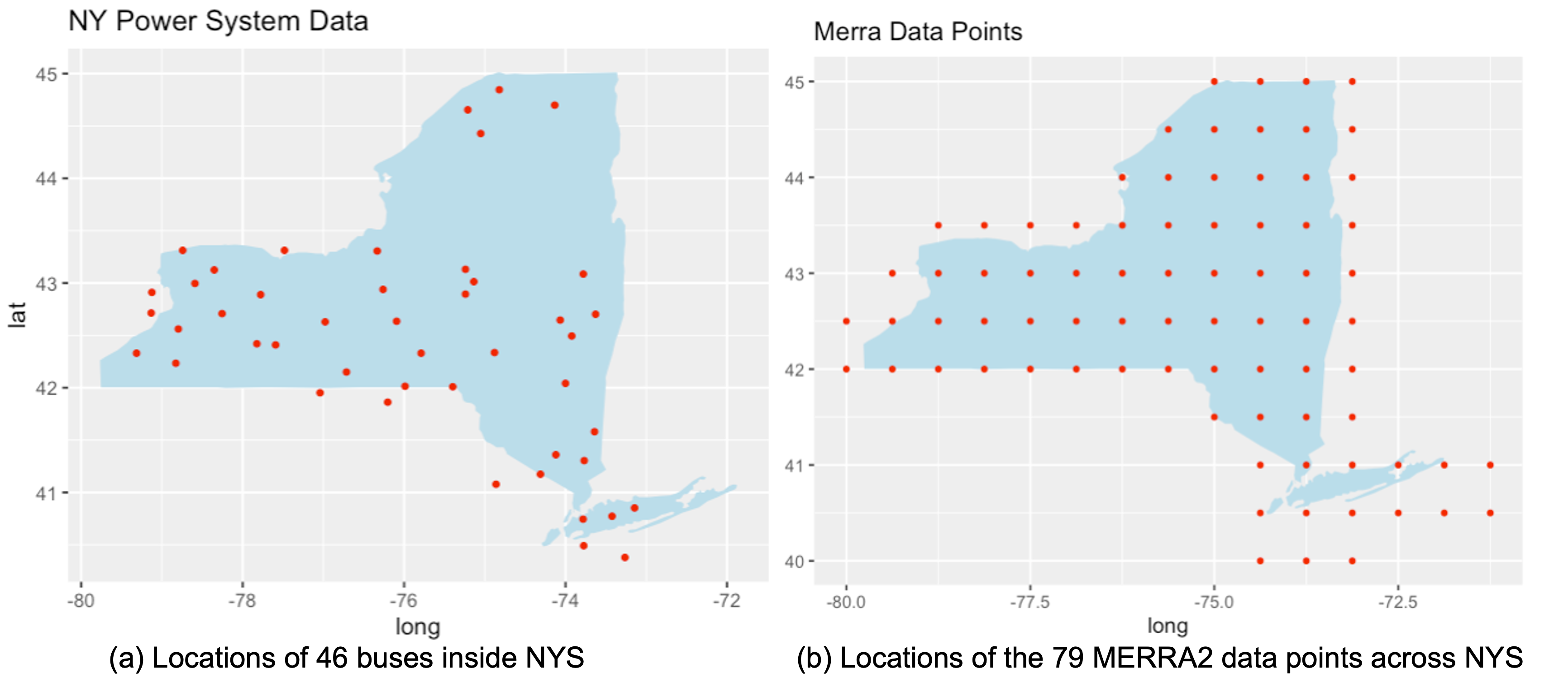}
\centering
\caption{Locations of the buses in the NYS power grid and locations of points for which we have MERRA-2 climate data}
\label{MERRA2DataPoints}
\end{figure}

We gather long records of grid-level wind speed across NYS from the MERRA-2 reanalysis product between 1980-2019. It includes hourly average wind speed at 10m surface level for 74 grid cells across NYS (see Figure \ref{MERRA2DataPoints}-b). We convert 10-m MERRA-2 wind speeds to power by applying the following steps at each land-based and offshore wind site: 
\begin{enumerate}
    \item We interpolate the 10m surface-level wind speed data of the closest grid to 100m level data using a power law.
    \item The interpolated wind speed data is then bias-corrected against the available NREL WTK wind speed data from 2007 to 2012 \cite{draxl2015wind}. Based on the associations between interpolated MERRA-2 wind speed data and NREL wind speed data, stability coefficients are calculated for each hour of each month of a year. The stability coefficients are used to bias correct the interpolated wind speed data from 1980 to 2019. For more detail on this step, see \cite{doering2023evaluating}.
    \item We model wind power generation based on the method presented in \cite{draxl2015wind}. Accordingly, wind speed data is converted to normalized wind power using the power curves. The normalized power is then scaled based on the capacity of the wind site.
\end{enumerate}

\textbf{Solar power}: We collect modeled solar site data from NREL Solar Integration National Dataset (SIND) \cite{bloom2016eastern}. The data includes the location of 6000 synthetic solar photovoltaic (PV) power plants over the United States and 5-minute solar power for all the simulated PV plants. Solar power plant locations are determined based on the capacity expansion plan for high-penetration renewables. The 5-minute solar power data is generated using the Sub-Hour Irradiance Algorithm \cite{NREL_solar}. Over NYS, the dataset includes 59 distributed-scale PV (DPV) solar sites with a total capacity of 1951 MW and 62 utility-scale PV (UPV) with a total capacity of 3798 MW. We pick all the simulated UPV solar sites and add them to our model. To meet the CLCPA targets for UPV power, we scale up the capacity of all UPV solar sites. The power generated by DPV solar sites is added to the model as a negative load. To meet the zonal capacities that we consider for our model, we add hypothetical DPV solar sites near populated areas in zone D, J and K. Figure \ref{windSolarLocations}-b represents the locations of all the above-mentioned solar sites. 

To simulate solar power at the above solar sites, we employ the method of statistical moments presented in \cite{perpinan2007calculation}. We collect long records of grid-level temperature and incident shortwave radiation from the MERRA-2 reanalysis data between 1980-2019. For each solar site, we estimate generated power using the closest grid cell's temperature and incident shortwave radiation. With a similar method that we used for wind power data, we bias-correct the estimated solar power data against the NREL data. Based on the associations between estimated solar power and available NREL solar power for 2006, stability coefficients are calculated for each month of the year. The stability coefficients are then used to bias correct the estimated solar power for all the years between 1980-2019. 

\textbf{Hydropower}: There exist 347 hydro plants across NYS \cite{liu2022open} where roughly 80$\%$ of the generated hydropower comes from the two largest hydropower plants: Robert Moses Niagara with a capacity of 2,675 MW and Moses-Saunders with a capacity of 856 MW. The power generated at Moses-Niagara is modeled using the model presented in \cite{meyer2017evaluating}. The NYS share of hydropower generated at the Moses-Saunders dam is estimated following the model developed in \cite{semmendinger2022establishing}. In both the Moses-Niagara and Moses-Saunders, the system's power production is simulated at a quarter-monthly time step. The quarter-monthly hydropower is disaggregated to an hourly timestep by assuming a constant amount of production for every hour of each quarter-month.

We aggregate the remaining 345 small hydro plants into eight hydropower plants. Table \ref{smallHydroCap} shows their capacity in the aggregated level, provided by the Gold Book \cite{GoldBook2019}. We assume a perfect correlation exists between all the small hydro plants and Moses Niagara. Therefore, their monthly output is proportional to their capacity. We estimate their monthly capacity factors by following these steps: 
\begin{enumerate}
    \item Using hourly hydropower generation from NYISO \cite{NYISOfuelMix}, the total amount of hydropower generation for each month is obtained.
    \item From the total monthly generation, Moses Niagara and Moses Saunders generated power is subtracted to obtain monthly generation by small hydropower plants.
    \item Dividing the monthly generation by the capacity of small hydro plants gives us their monthly capacity factors, summarized in Table \ref{smallHydroCapfactor}.
\end{enumerate}
Similar to the two big hydropower plants, we assume that the amount of production for every hour of each month is constant and accordingly disaggregate the monthly hydropower to hourly timestep. The power generated by small hydro plants is added to the model as a negative load. 

\renewcommand{\thetable}{A\arabic{table}}
\setcounter{table}{0}

\begin{table}[!b]
\centering
\caption{Aggregated capacity of small hydropower plants}
\begin{tabular}{|lll|}
\hline
Bus index & Capacity & Zone \\ \hline
53        & 63.8     & B    \\
50        & 71.4     & C    \\
68        & 38       & C    \\
48        & 58.8     & D    \\
45        & 28.6     & E    \\
47        & 347.7    & E    \\
42        & 269.6    & F    \\
77        & 75.8     & G    \\ \hline
total     & 953.7    &      \\ \hline
\end{tabular}
\label{smallHydroCap}
\end{table}
\begin{table}[!t]
\centering
\caption{Monthly capacity factor of small hydropower plants}
\begin{tabular}{|ll|ll|}
\hline
Month 	  & Capacity factor & Month 	  & Capacity factor\\ \hline
1         & 0.576    & 7         & 0.388 \\
2         & 0.551    & 8         & 0.328 \\
3         & 0.642    & 9         & 0.278 \\
4         & 0.663    & 10        & 0.371 \\
5         & 0.567    & 11        & 0.523 \\
6         & 0.397    & 12        & 0.564 \\ \hline
\end{tabular}
\label{smallHydroCapfactor}
\end{table}

\begin{figure}[!b]
\includegraphics[width=\textwidth]{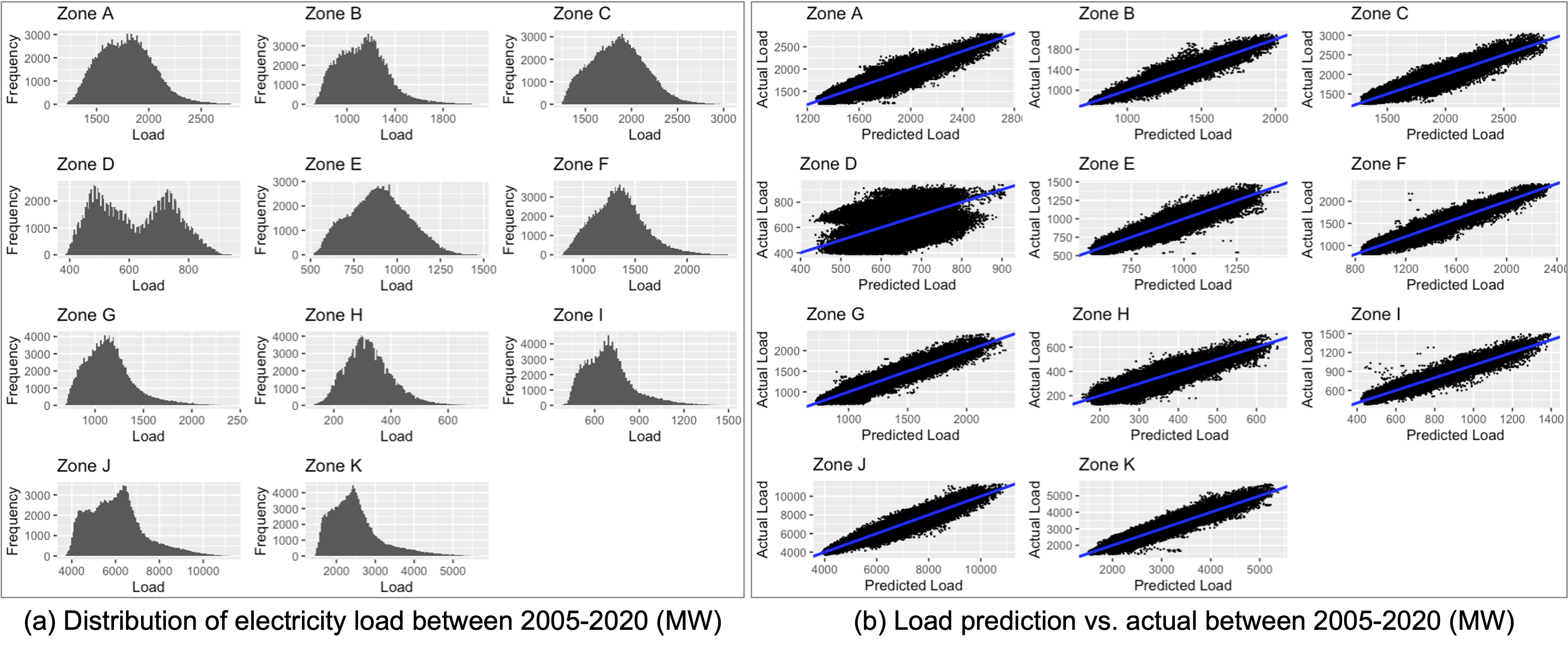}
\centering
\caption{(a) shows the distribution of actual electricity load (MW) per zone, and (b) shows the scatterplot of out-of-sample predictions versus actual electricity load for each zone. The blue line is the identity line which represents the goodness of fit.}
\label{loadData}
\end{figure}

\textbf{Energy load}: We collect historical hourly load data between 2005-2019 for all NYS load zones from NYISO \cite{NYISOload}. In order to extend the hourly load record for the remaining years between 1980-2004, we develop a random forest (RF) model for each load zone using the actual data. The model is selected from among other models due to its high out-of-sample predictive performance. In the developed models, for each load zone, feature vectors include the last 24 hours of temperature, day of the week, and day of the year, and the target variable is hourly energy load. The hourly air temperature is averaged across each zone from the MERRA-2 reanalysis product. Once the model is fit, to ensure the variance of the predicted load is not underestimated by the model, predictions for years 1980-2004 are bias corrected using the empirical quantile mapping approach according to the following steps:
\begin{enumerate}
    \item CDFs of historical observations and predicted values (for data between 2005-2019) are first estimated over a set of regularly spaced quantile levels, $\tau =  0, 0.005, 0.01, …, 0.995, 1.00$. 
    \item The corresponding quantile for each predicted value (data between 1980-2004) is estimated using the CDF of predicted values (from step 1).
    \item The obtained quantile is mapped into the corresponding values from the CDFs of historical data (from step 1).  
\end{enumerate}
The bias-corrected data is then combined with the actual data to construct a complete record of loads for the study period of 1980-2019. All load data is then scaled up to arrive at the summer and winter peak load levels forecasted by NYISO \cite{GoldBook2021} for the year 2030 under the CLCPA case. In the estimation of the peak loads that we employed, impacts of energy efficiency programs, electrification, electric vehicle usage, and other load modifying impacts are considered. The acquired summer and winter Peak loads of the New York control area for years 2020 and 2030 are shown in Table \ref{CLCPApeakLoad}. To scale up the load, we increase the loads over months April-September by $4\%$, and loads of months October-March by $3\%$. Zonal hourly load data are disaggregated and divided between buses according to the original load ratio of the NPCC-140 model \cite{chow1991user}.

\begin{figure}[!t]
\includegraphics[width=\textwidth]{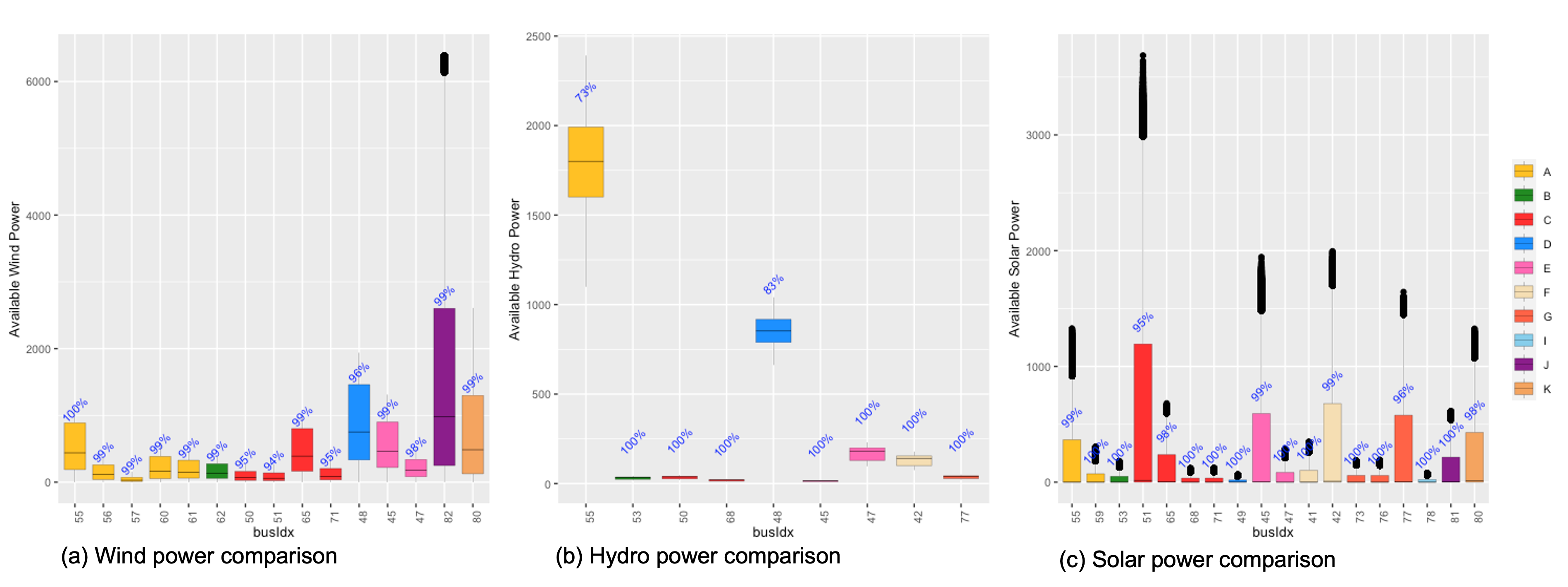}
\centering
\caption{40-year average of renewable power generation (MW) per bus. The value on top of each box is the percentage of average dispatched power.}
\label{busLevelRES}
\end{figure}
\begin{table}[!b]
\centering
\caption{Generated power (MW) average and deviation ($\%$) from average in various temporal resolutions. For thermal and nuclear power, the reported values are the total nameplate capacity of corresponding generators.}
\label{Powerdev_generated}
\begin{tabular}{|l|llll|}
\hline
                   & Yearly         & Monthly           & Daily             & Hourly    \\ \hline
Wind    & 5.5e+07 ±8.8\%   & 4.6e+06 ±81\% 	& 1.5e+05 ±184\% 	& 6.3e+03 ±185\% \\
Solar   & 2.3e+07 ±7.0\%   & 1.9e+06 ±62\% 	& 0.6e+05 ±106\% 	& 2.6e+03 ±474\% \\
hydro   & 2.7e+07 ±17.2\%  & 2.3e+06 ±28\% 	& 0.7e+05 ±32\% 	& 3.1e+03 ±32\%  \\
wind-solar-hydro   & 10.1e+07 ±8.3\%  & 8.7e+06 ±38\% 	& 2.9e+05 ±103\% 	& 11.9e+03 ±197\%  \\
Thermal & 25.0e+07       & 20.9e+06          & 6.9e+05  	        & 28.6e+03 	\\
Nuclear & 2.9e+07        & 2.4e+06 	        & 0.8e+05  	        & 3.4e+03  	\\ \hline
\end{tabular}
\end{table}
\subsection*{Results} \label{AppendixResults}  
Table \ref{Powerdev_generated} summarizes the power generation (MW) average and deviation ($\%$) from long-term averages in various timescales and for different sources. Hydropower is considerably more volatile than the other two renewable sources in a yearly timescale. conversely, it is the most stable renewable power source in smaller timescales, including monthly, daily, and hourly. 

Figure \ref{busLevelRES} represents the long-term average of wind, solar, and hydro power generation per bus. Off-shore wind power concentrated in zones J and K (buses 82 and 80) is more variable than land-based wind power. The Moses Niagara generation is more volatile than the Moses Saunders generation and has more curtailment. Solar power is distributed more evenly across the state than wind and hydropower. However, at the bus level, solar generation of buses with more UPV sources is more volatile than buses connected to DPV.  

\begin{figure}[t]
\includegraphics[width=\textwidth]{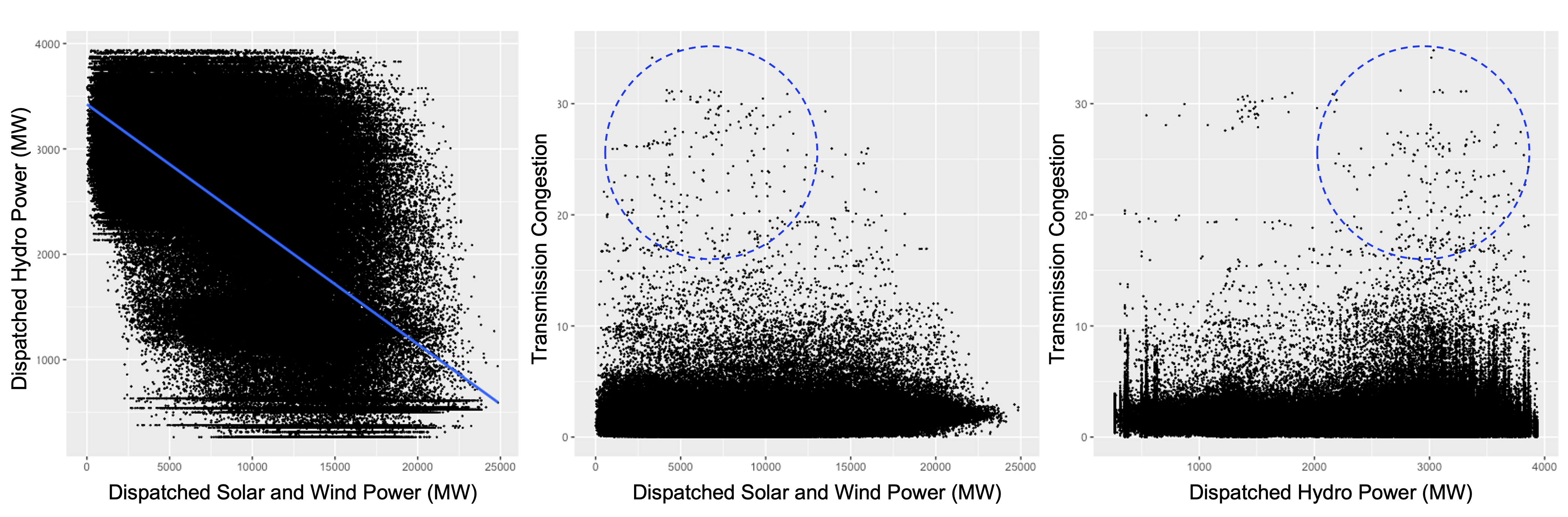}
\centering
\caption{Scatterplots of dispatched renewables and their relationship with transmission congestion}
\label{dispatchedVScongestion}
\end{figure}

\begin{figure}[t]
\includegraphics[width=\textwidth]{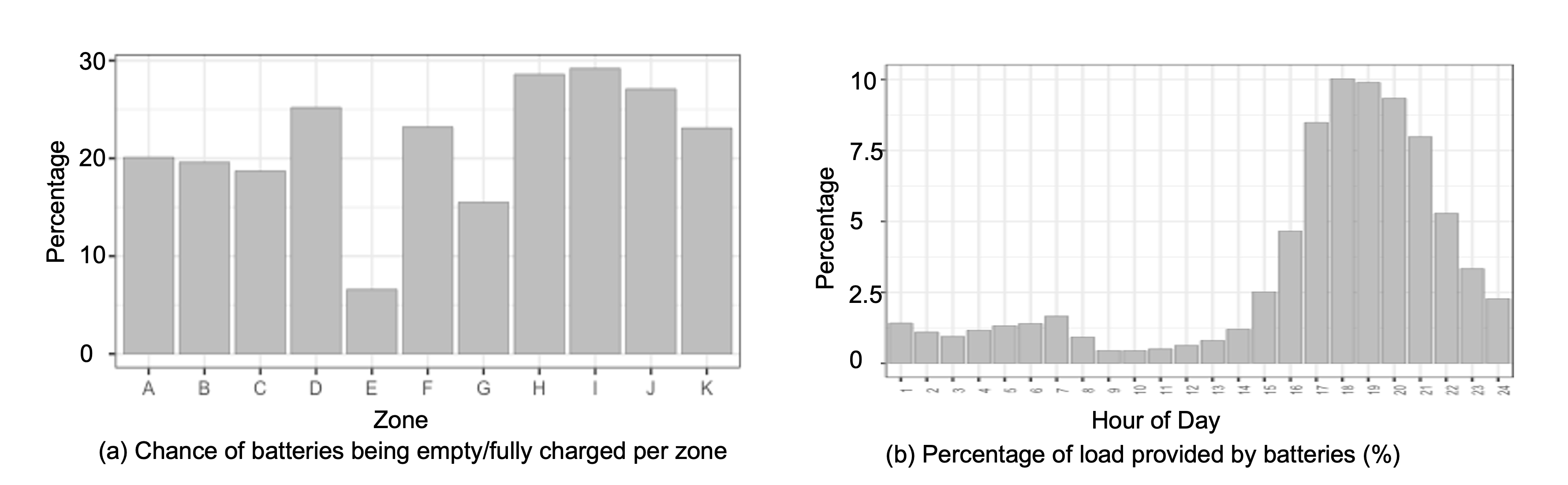}
\centering
\caption{Comparing the percentage of average battery usage in (a) different zones and (b) different hours of a day.}
\label{batteryUsage} 
\end{figure}

According to Figure \ref{dispatchedVScongestion}, dispatched hydropower increases as dispatched wind and solar power decrease. The transmission system is highly congested during the low amount of dispatched wind and solar power and high dispatched hydropower. Finally, Figure \ref{batteryUsage} summarizes the percentage of average battery usage in different zones and different hours of the day. If batteries are frequently fully charged, there is more generation than consumption, so extra storage capacity can help reduce curtailment. Conversely, if batteries are mostly empty, this shows more consumption than generations. Further storage capacity, in this case, help to avoid power shortage. Therefore, Figure \ref{batteryUsage}-a represents the likely need for extra batteries per zone. Figure \ref{batteryUsage}-b, on the other side, represents in which hours of the day batteries are mostly used as the power source. 
\subsection*{Sensitivity Analysis} \label{AppendixSensitivityAnal}
For sensitivity analysis, we focus on year 32, a stressful year with low renewable generation, high average LMP, and a large number of price spikes. We evaluate the sensitivity of state-level wind and solar generation and average LMPs under multiple scenarios of (i) renewable and storage capacity allocations within zones while meeting zonal capacities suggested by the Congestion Assessment and Resource Integration Study \cite{CARIS2019}, and (ii) zonal-level renewable capacities while meeting CLCPA requirements.

\begin{figure}[!t]
\includegraphics[width=\textwidth]{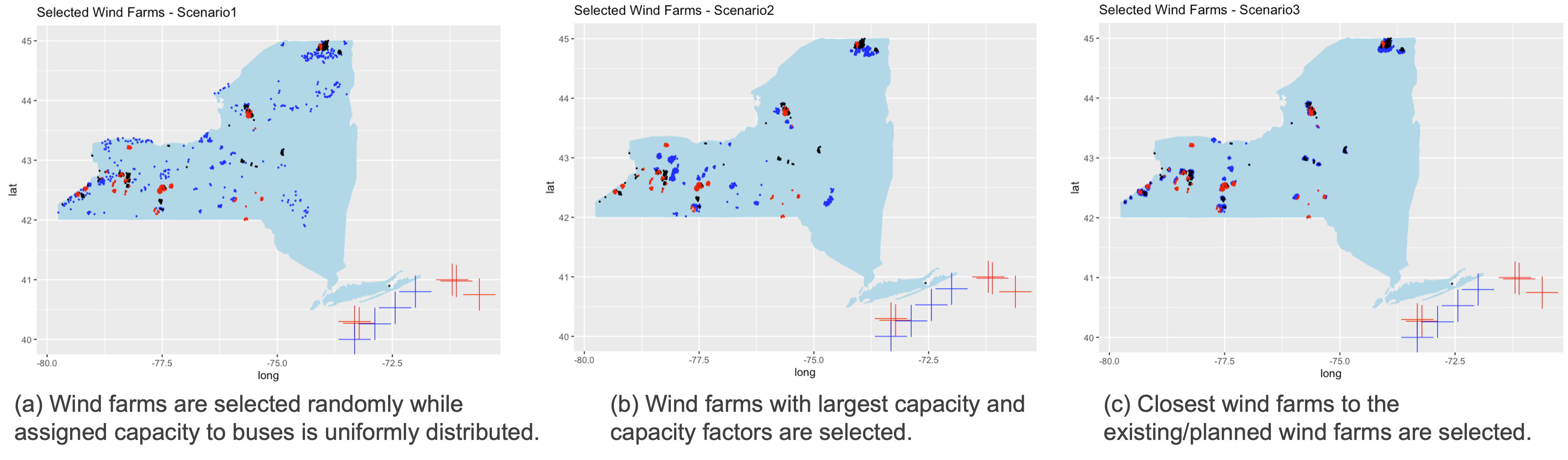}
\centering
\caption{Different types of wind farm selection}
\label{windCapScen} 
\end{figure}

\begin{figure}[!b]
\includegraphics[width=\textwidth]{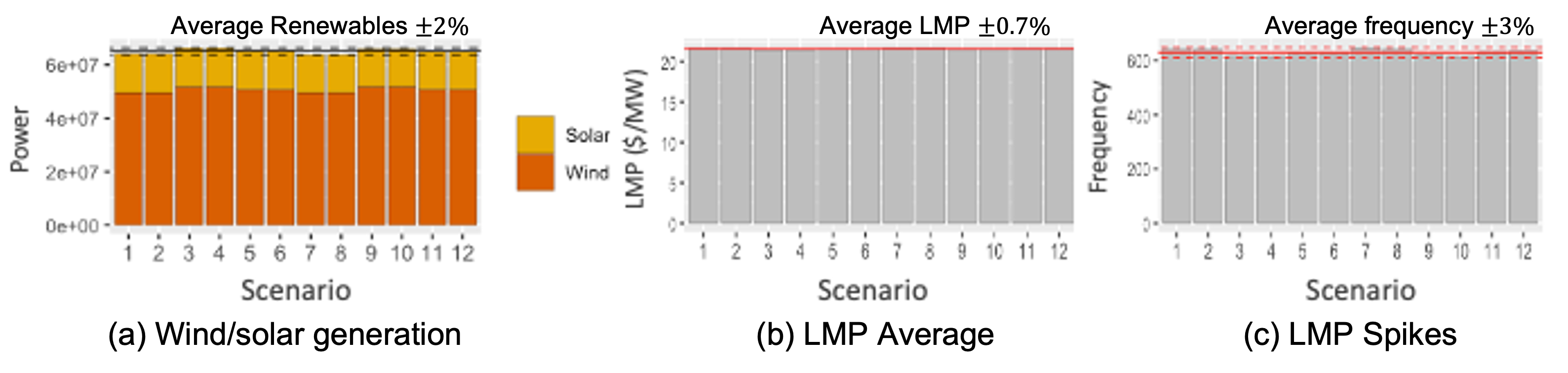}
\centering
\caption{Comparing renewable generation and LMPs under different scenarios of wind turbine and solar site capacity and allocation decisions}
\label{SensAnal1} 
\end{figure}

For sensitivity analysis of renewable and storage capacity allocation decisions, we consider 12 different scenarios. They include three types of wind farm selection, two types of solar site capacity allocations, and two types of storage capacity assignments. Wind farm selection cases include (1) random selection, (2) selection of wind farms with the largest capacity and capacity factors and (3) selection of wind farms that are closest to existing farms. Figure \ref{windCapScen} represents the distribution of wind farms under each case. Solar site capacity assignment is also based on (1) dispersed and (2) concentrated strategies. In the former case, we equally divide NREL zonal solar capacity between buses, and then we scale up the capacity of each bus to meet CARIS zonal capacities. In the latter case, we reverse the order, scale up the capacities first, and then divide them between buses. Finally, for storage units, we considered disaggregating zonal capacities into bus-level capacities based on (1) the load proportion of each bus and (2) the total renewable generation of each bus. Results show that our decisions regarding bus-level renewable and storage capacity allocation do not have considerable effects on the renewable generation and LMPs (see Figure \ref{SensAnal1}).


\begin{figure}[t]
\includegraphics[width=\textwidth]{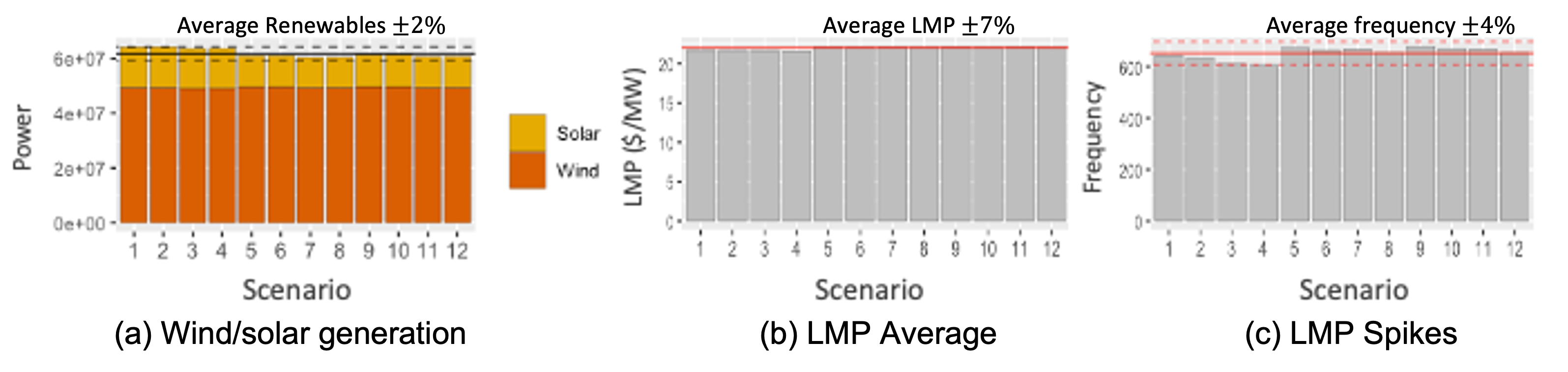}
\centering
\caption{Comparing renewable generation and LMPs under different scenarios of wind turbine and solar site capacity and allocation decisions}
\label{SensAnal2} 
\end{figure}

For analyzing the sensitivity of our results to zonal-level renewable capacities suggested by CARIS \cite{CARIS2019}, we develop 12 different scenarios. In these scenarios, the zonal-level capacities have changed while state-level capacities are unchanged, as targeted by CLCPA. To disaggregate to bus level, we employed the base strategy described in the main text. The 12 scenarios result from two wind capacity plans, three solar capacity plans, and two storage capacity plans, as summarized in Table \ref{SensAnal2_cap}. Results show that annual renewable generation and average price still do not change substantially, while the frequency of price spikes by $7\%$ (see Figure \ref{SensAnal2}). 

\begin{table}[!b]
\centering
\caption{Zonal capacities of wind, solar, and storage units under different test cases}
\label{SensAnal2_cap}
\footnotesize
\begin{tabular}{|c|cc|cc|cc|cc|cc|cc|}
\hline
\multicolumn{1}{|l|}{\multirow{2}{*}{Zone}} & \multicolumn{2}{c|}{Wind Cap1}                                                                               & \multicolumn{2}{c|}{Wind Cap2}                                                                               & \multicolumn{2}{c|}{Solar Cap1} & \multicolumn{2}{c|}{Solar Cap2} & \multicolumn{2}{c|}{Solar Cap3} & \multirow{2}{*}{\begin{tabular}[c]{@{}c@{}}Storage \\ Cap1\end{tabular}} & \multirow{2}{*}{\begin{tabular}[c]{@{}c@{}}Storage \\ Cap2\end{tabular}} \\ \cline{2-11}
\multicolumn{1}{|l|}{}                      & \begin{tabular}[c]{@{}c@{}}Land-\\ based\end{tabular} & \begin{tabular}[c]{@{}c@{}}Off-\\ shore\end{tabular} & \begin{tabular}[c]{@{}c@{}}Land-\\ based\end{tabular} & \begin{tabular}[c]{@{}c@{}}Off-\\ shore\end{tabular} & UPV            & DPV            & UPV            & DPV            & UPV            & DPV            &                                                                          &                                                                          \\ \hline
A                                           & 2692                                                  &                                                      & 3085                                                  &                                                      & 704            & 5748           & 704            & 7091           & 704            & 6141           & 150                                                                      & 223                                                                      \\
B                                           & 390                                                   &                                                      & 783                                                   &                                                      & 218            & 656            & 218            & 2000           & 218            & 1049           & 90                                                                       & 164                                                                      \\
C                                           & 1923                                                  &                                                      & 2316                                                  &                                                      & 596            & 3585           & 596            & 4928           & 596            & 3978           & 120                                                                      & 193                                                                      \\
D                                           & 1935                                                  &                                                      & 756                                                   &                                                      & 69             &                & 69             &                & 69             &                & 180                                                                      & 100                                                                      \\
E                                           & 1821                                                  &                                                      & 1821                                                  &                                                      & 673            & 2268           & 673            & 2268           & 673            & 2268           & 120                                                                      & 120                                                                      \\
F                                           &                                                       &                                                      &                                                       &                                                      & 827            & 4661           & 827            & 631            & 827            & 3482           & 240                                                                      & 100                                                                      \\
G                                           &                                                       &                                                      &                                                       &                                                      & 684            & 2636           & 684            & 2636           & 684            & 2636           & 100                                                                      & 100                                                                      \\
H                                           &                                                       &                                                      &                                                       &                                                      & 61             &                & 61             &                & 61             &                & 100                                                                      & 100                                                                      \\
I                                           &                                                       &                                                      &                                                       &                                                      & 90             &                & 90             &                & 90             &                & 100                                                                      & 100                                                                      \\
J                                           &                                                       & 6391                                                 &                                                       & 6391                                                 & 672            &                & 672            &                & 672            &                & 150                                                                      & 150                                                                      \\
K                                           &                                                       & 2609                                                 &                                                       & 2609                                                 & 846            & 77             & 846            & 77             & 846            & 77             & 480                                                                      & 480                                                                      \\ \hline
Total                                       & 8761                                                  & 9000                                                 & 8761                                                  & 9000                                                 & 5439           & 19631          & 5439           & 19631          & 5439           & 19631          & 5439                                                                     & 5439                                                                     \\ \hline
\end{tabular}
\end{table}

We further investigated the impacts of thermal generator retirement on load shedding under two different scenarios. In the first scenario, the retired capacity is divided uniformly between the zones and buses. In the second scenario, however, we initially retire generators of zone A-F (upstate zones); if extra retirement is needed, we retire generators from zones G-K (downstate zones). The total capacity of retired thermal generators is the same in both scenarios. Results show that both scenarios result in a large load shedding (see Table \ref{thermalGenretirement}).

\begin{table}[!t]
\centering
\caption{Comparing two different scenarios of thermal generator retirement plans}
\label{thermalGenretirement}
\footnotesize
\begin{tabular}{|cc|ll|cc|}
\hline
\multicolumn{2}{|c|}{Retired thermal power (MW)}  & \multicolumn{2}{c|}{Load shedding over the year}                                                                                                                                                                     & \multicolumn{2}{c|}{Frequency of nonzero load shedding}      \\ \hline
\multicolumn{1}{|c|}{Inside NY} & Outside NY & \multicolumn{1}{c|}{Retirement Scen. 1}                                                                             & Retirement Scen. 2                                                                             & \multicolumn{1}{c|}{Retirement Scen. 1} & Retirement Scen. 2 \\ \hline
\multicolumn{1}{|c|}{0}         & 0          & \multicolumn{1}{c|}{0}                                                                                              & 0                                                                                              & \multicolumn{1}{c|}{0}                  & 0                  \\
\multicolumn{1}{|c|}{2706}      & 7018       & \multicolumn{1}{c|}{0}                                                                                              & 0                                                                                              & \multicolumn{1}{c|}{0}                  & 0                  \\
\multicolumn{1}{|c|}{5412}      & 14037      & \multicolumn{1}{c|}{0}                                                                                              & 0                                                                                              & \multicolumn{1}{c|}{0}                  & 0                  \\
\multicolumn{1}{|c|}{8119}      & 21056      & \multicolumn{1}{c|}{\begin{tabular}[c]{@{}c@{}}0-7.7\% of load\\ Ave:0.006\% of load\\ Total:14543 MW\end{tabular}} & \begin{tabular}[c]{@{}c@{}}0-7.7\% of load\\ Ave:0.006\% of load\\ Total:14543 MW\end{tabular} & \multicolumn{1}{c|}{17}                 & 0                  \\
\multicolumn{1}{|c|}{10825}     & 28075      & \multicolumn{1}{c|}{\begin{tabular}[c]{@{}c@{}}0-25\% of load\\ Ave:0.13\% of load\\ Total:286151 MW\end{tabular}}  & \begin{tabular}[c]{@{}c@{}}0-25\% of load\\ Ave:0.15\% of load\\ Total:324583 MW\end{tabular}  & \multicolumn{1}{c|}{184}                & 266                \\
\multicolumn{1}{|c|}{13532}     & 35094      & \multicolumn{1}{c|}{\begin{tabular}[c]{@{}c@{}}0-42\% of load\\ Ave:1.3\% of load\\ Total:2579146 MW\end{tabular}}  & \begin{tabular}[c]{@{}c@{}}0-45\% of load\\ Ave:1.3\% of load\\ Total:2580140 MW\end{tabular}  & \multicolumn{1}{c|}{1115}               & 1172               \\
\multicolumn{1}{|c|}{16238}     & 42113      & \multicolumn{1}{c|}{\begin{tabular}[c]{@{}c@{}}0-58\% of load\\ Ave:5.8\% of load\\ Total:11036222 MW\end{tabular}} & \begin{tabular}[c]{@{}c@{}}0-59\% of load\\ Ave:6.1\% of load\\ Total:11373952 MW\end{tabular} & \multicolumn{1}{c|}{3395}               & 3048               \\ \hline
\end{tabular}
\end{table}

\end{onehalfspacing}
\end{document}